\documentstyle[12pt]{article}

\newtheorem{prop}{Proposition}
\newtheorem{deff}{Definition}
\newtheorem{corr}{Corollary}

\title{\Huge The Wigner-Weyl-Moyal Formalism on Algebraic Structures}
\author{\large Frank Antonsen \\University of Copenhagen \\ Niels Bohr Institute}

\begin{document}
\maketitle

\begin{abstract}
We first introduce the Wigner-Weyl-Moyal formalism for a theory whose 
phase-space is
an arbitrary Lie algebra. We also generalize to quantum Lie algebras and to
supersymmetric theories. It turns out that the non-commutativity leads to a
deformation of the classical phase-space: instead of being a vector space it
becomes a manifold, the topology of which is given by the commutator
relations. It is shown in fact that the classical phase-space, for a semi-simple
Lie algebra, becomes a
homogenous symplectic manifold. The symplectic product is also deformed. 
We finally make some comments on how to generalize to $C^*$-algebras and 
other operator algebras too.
\end{abstract}

\section{Introduction}
The very powerful Wigner-Weyl-Moyal (WWM) formalism 
\cite{Grossman,Royer,Dahl,su2,Li} is a way to associate with
each operator describing a state, observable or transition, a function on phase
space. This function is known as the Weyl symbol, or the Weyl transform of the
corresponding operator. In this way the wave function 
(or rather the density matrix) is
associated with a pseudo-distribution function known as the Wigner function.
This function, denote it by $F$, is the closest analogue of the classical
phase-space distribution, which enters for instance in the Boltzmann equation.
It can, however, be non-positive, and is hence not a proper distribution
function -- in most cases the Heisenberg uncertainty relations forbids the
existence of such a proper distribution function. As first pointed out by 
Moyal, the Weyl transform generates a deformation, on the phase-space, of the 
classical
Poisson brackets and of the usual commutative product, $(f(q,p),g(q,p))
\rightarrow f(q,p)g(q,p) = (fg)(q,p)$. The deformed product is denoted by $*$
and is called the twisted product. It is in general non-commutative. The
deformation of the Poisson bracket is what is known as the Moyal bracket
\begin{displaymath}
    [f(q,p),g(q,p)]_M = f(q,p)*g(q,p)-g(q,p)*f(q,p) = i\hbar \{f(q,p),g(q,p)
    \}_{PB} + O(\hbar^2)
\end{displaymath}
It is this method we want to extend to a phase-space which is not just that of
quantum mechanics, but can be an arbitrary (finite or infinite dimensional) Lie
algebra or, as will be shown later, a super-Lie algebra, a quantum-Lie
algebra or a $C^*$-algebra.\footnote{Some abuse of notation is used 
here. When we say that a
quantum mechanical phase-space is given by (or simply {\em is}) some Lie
algebra, what we mean is that any quantum physical observable is some function
of the generators of this algebra, hence the quantum phase space is really the
universal enveloping algebra, $U$, of the Lie algebra in question. It is, 
however, straightforward to go from the Lie algebra to its universal enveloping 
algebra -- the algebra of formal power series with elements from the Lie 
algebra. 
Furthermore, one could just aswell consider the skew field, $P$, of fractions
of $U$, $P=\{u^{-1}v~|~u,v\in U\}$. This would correspond to an algebra of
formal Laurent series (i.e. functions possibly with singularities), and the 
correspondinig classical phase space would then consists of meromorphic 
functions.}\\
We will first review the standard WWM approach to the quantum mechanical
phase-space, i.e. to the Lie algebra, $h_n$, of the Heisenberg group in $n$
dimensions. This
will be done in terms of certain translation operators. This formalism will 
then
be carried over into a second quantized formulation by introducing a new basis,
namely that of creation and annihilation operators. This will at once show us
how to extend the formalism in two directions: (1) to an arbitrary Lie algebra,
and (2) to fermionic degrees of freedom. These can then be combined to give a
WWM formalism for super-Lie algebras. The way we derive the standard WWM
approach will show some connection with quantum groups, and hence we will also 
be commenting on how to extend this formalism even further, into the realm of
quantum deformed Lie algebras -- quantum-Lie algebras. Finally we will study
general operator algebras, and we will show that our method can be 
generalized to $C^*$-algebras. We finish off with some comments on further
generalizations and applications.

\section{The WWM Approach to the Standard Phase-Space}
The standard phase-space of quantum mechanics is given by $2n$ generators 
$\hat{q}_i, \hat{p}_i$ satisfying (we'll only treat bosons for now, we will, 
however, return to fermions later)
\begin{equation}
    [\hat{q}_i, \hat{p}_j ] = i\delta_{ij}\mbox{    with }i,j=1,...,n
\end{equation}
in units where $\hbar=1$.\\
We know that these commutation relations can only be represented faithfully in
terms of operators on some Hilbert space, leading to the standard formulation 
of
quantum theory. We're interested in a phase-space formulation which as closely
as possible resembles that of classical stastical mechanics, and we thus need a
correspondence between observables represented by operators on the 
Hilbert space
$H=L^2(X)$ ($X$ is the coordinate space, $q$-space, i.e. an $n$ dimensional
vector space) and functions on a $2n$-dimensional symplectic space, 
phase-space,
i.e. we want a map, the {\em Weyl map}, $\hat{A}\mapsto A_W(q,p)$, where 
$\hat{A}$ is an operator on 
$H$ and $A_W$ is some function on the classical phase-space. Quantization as 
a general formalism related to the introduction of such symbols for operators 
was first extensively studied by Berezin, I think, \cite{Berezin}. Following 
Grossmann, Royer and Dahl, \cite{Grossman,Royer,Dahl} (see also Li 
\cite{Li}), we introduce operators
\begin{equation}
    \Pi(u,v) = \exp(i(u\cdot\hat{p}-v\cdot\hat{q}))
\end{equation}
these satisfy
\begin{equation}
    \Pi(u,v)\Pi(u',v') = \Pi(u+u',v+v')Q(u,v;u',v')
\end{equation}
where
\begin{equation}
    Q(u,v;u',v') = e^{i\frac{1}{2}(uv'-vu')}
\end{equation}
is a C-number function. This shows then that $\Pi(u,v)$ constitutes a ray
representation of the Euclidean group ${\bf {\sf R}}^{2n}$, the group of translations 
in the Euclidean plane.\footnote{I use the following notation for the most 
important sets of numbers: ${\bf {\sf N}}$ is the natural numbers, ${\bf {\sf N}}=\{1,2,...\}$, ${\bf {\sf Z}}$ 
denotes the integers, ${\bf {\sf Q}}$ the rationals, ${\bf {\sf R}}$ the reals, ${\bf {\sf C}}$ the complex 
numbers and ${\bf {\sf H}}$ the quaternions. A general field (or even division ring) 
will be denoted by ${\bf {\sf F}}$, while ${\bf {\sf T}}$ denotes the torus, ${\bf {\sf T}}=\{z\in{\bf {\sf C}}~|~ 
|z|=1\}\simeq S^1$.\\
Commutators and anticommutators will be denoted by $[\cdot,\cdot],$ and 
$\{\cdot,\cdot\}$, while Moyal and Poisson brackets will be characterized 
by subscripts $M$ and $PB$ respectively.} 
One easily proves
\begin{eqnarray}
    \Pi(u,v)\hat{p}\Pi(u,v)^{-1} &=& \hat{p}-v\\
    \Pi(u,v)\hat{q}\Pi(u,v)^{-1} &=& \hat{q}-u
\end{eqnarray}
which gives us a physical picture of what these operators do: they are
translations in phase-space. It also shows us that $u$
acts like a C-number version of the Q-number $\hat{q}$ and $v$ as a C-number 
version of the Q-number $\hat{p}$, this shows that $\{(u,v)\}$ can be 
identified with the {\em classical} phase-space. There are no restrictions
imposed upon $u,v$, hence the classical phase-space becomes simply ${\bf {\sf R}}^{2n}$.\\
We can use the operator $\Pi(u,v)$ to construct our map $\hat{A}\mapsto 
A_W(u,v)$ as follows. To each operator describing an observable we associate 
a function given by
\begin{equation}
    A_W(u,v) = {\rm Tr}(\Pi(u,v)\hat{A})
\end{equation}
this can be inverted to give
\begin{equation}
    \hat{A} = \int A_W(u,v)\Pi(u,v) dudv
\end{equation}
Actually, this map is only an isomorphism when $\hat{A}$ lies in the space
${\cal B}^2(H)$ of Hilbert-Schmidt operators. And we thus have an isomorphism
between the space of Hilbert-Schmidt operators on $L^2({\bf {\sf R}}^n)$ and the function
space $L^2({\bf {\sf R}}^n\times {\bf {\sf R}}^n)$. The function corresponding to the density matrix
$\rho$ is known as the {\em Wigner function} (strictly speaking this is only 
the symplectic Fourier transform of the proper Wigner function). 
For a pure state $\psi$ we have $\rho = |\psi\rangle\langle\psi|$ and hence
\begin{equation}
    F(u,v) = {\rm Tr}(\Pi(u,v)|\psi\rangle\langle\psi|) = \langle\psi|\Pi(u,v)
    |\psi\rangle
\end{equation}
which gives a geometric interpretation of the Wigner function: it is the
expectation value of a reflection operator (the symplectic Fourier transform 
of the translation operator $\Pi$ is a reflection operator). 
This Wigner function is the closest
quantum cousin of the classical distribution function $f(q,p)$, it is, however,
in general non-positive.\\
The {\em Weyl-map} $\hat{A}\mapsto A_W$ generates an algebra structure on
$L^2({\bf {\sf R}} \times {\bf {\sf R}})$ via
\begin{equation}
    (\hat{A}\hat{B})_W \equiv A_W*B_W
\end{equation}
This product is known as the {\em twisted product}, it is non-commutative but
associative, hence with this product $L^2({\bf {\sf R}}\times {\bf {\sf R}})$ becomes a non-abelian
Banach algebra (a Hilbert-algebra even). One can show\footnote{A few papers have
been written in the mathematics literature dealing with twisted products for
some classical groups, see e.g. \cite{Moreno}.}
\begin{equation}
    f*g = f(u,v)\exp(-\frac{1}{2}i\hbar\frac{\partial}{\partial v}\cdot
    \frac{\partial}{\partial u})g(u,v)
\end{equation}
where $\partial/\partial v$ is understood always to act on $f(u,v)$ and the
other derivative always to act on $g$. We have reinserted $\hbar$ for 
clarity.\\
As the twisted product is non-commutative we can introduce a kind of 
commutator, known as the {\em Moyal bracket}
\begin{equation}
    [f(u,v),g(u,v)]_M \equiv f*g-g*f
\end{equation}
One easily sees that
\begin{equation}
    \left([\hat{A},\hat{B}]\right)_W = [A_W,B_W]_M
\end{equation}
Furthermore
\begin{equation}
    [f,g]_M = 2if\sin(\frac{1}{2}\hbar\bigtriangleup)g
\end{equation}
where we have introduced the {\em bi-differential operator}
\begin{equation}
    f\bigtriangleup g \equiv \frac{\partial f}{\partial v}\cdot\frac{\partial g}
      {\partial u}- (u\leftrightarrow v) = \{f,g\}_{\rm PB}
\end{equation}
which is the bi-differential operator defining the classical Poisson brackets,
$\{\cdot,\cdot\}_{\rm PB}$. Hence
\begin{equation}
    \left([\hat{A},\hat{B}]\right)_W = [A_W,B_W]_M =i\hbar\{A_W, B_W\}_{\rm PB}
    + O(\hbar^2)
\end{equation}
thus the Moyal bracket is a deformation of the classical Poisson bracket. Such
deformations of classical Poisson structures have also been studied in their own
right in the mathematics literature, I refer to \cite{EK}. Also
note that this relation clarifies the usual Heisenberg quantization rule
\begin{displaymath}
    \{\cdot,\cdot\}_{\rm PB} \rightarrow \frac{1}{i\hbar}[\cdot,\cdot]
\end{displaymath}
One should note that the Wigner function considered as a mapping ${\cal B}^2
\rightarrow L^2({\bf {\sf R}}^{2n})$ is not unique, one can modify the definition by the
inclusion of an arbitrary function, see Cohen \cite{Cohen}.
Each such function corresponds to a
different prescription for the ordering of operator products. The Wigner
function is, however, the simplest of these functions, and the only one for
which we do not need a ``dual'' for going the other way $L^2({\bf {\sf R}}^{2n})
\rightarrow {\cal B}^2$. I refer to \cite{Dahl2,Cohen} for 
further details.\\
Furthermore, one could just aswell use a translation operator based on {\em 
all} the generators of the Lie algebra, i.e. using
\begin{displaymath}
    \Pi_{\rm alt}(u,v,w) \equiv \exp(iu\hat{p}-iv\hat{q}+iw\hat{1}) 
\end{displaymath}
and the classical ``phase-space'' is now apparently three-dimensional
(parametrized by $u,v,w$), but one should note that $\hat{1}$ lies in the 
center
of the algebra (the Heisenberg algebra is a central extension of the algebra of
translations ${\bf {\sf R}}^2$), hence including it simply amounts to multiplying the 
functions by a phase:
\begin{displaymath}
    \Pi_{\rm alt}(u,v,w) = e^{iw}\Pi(u,v)
\end{displaymath}
and can thus be ignored. These comments will turn out to be useful when the
generalization to arbitrary Lie algebras is attempted.\\
Fascinating as all this is we nonetheless have to move on. We want to 
generalize
the above outlined beautiful formalism to the case where the phase-space is not
just the Heisenberg algebra $h_n$, but any Lie algebra $\bf g$.

\subsection{Creation and Annihilation Operators}
We need one more step, before we can safely generalize to arbitrary Lie
algebras. All physical processes can be described in terms of creation and
annihilation operators. For a simple (bosonic) quantum mechanical system we 
know that these are given in terms of the operators $\hat{q},\hat{p}$ by
\begin{eqnarray}
    \hat{a} &=& \frac{1}{\sqrt{2}}\left(\hat{p}+i\hat{q}\right)\\
    \hat{a}^\dagger &=& \frac{1}{\sqrt{2}}\left(\hat{p}-i\hat{q}\right)
\end{eqnarray}
i.e. by a simple rotation of the quantum phase-space. We know that these
operators satisfy
\begin{eqnarray}
    \left[\hat{a},\hat{a}^\dagger\right] &=& 1\\
    \left[\hat{n},\hat{a}\right] &=& -\hat{a}\\
    \left[\hat{n},\hat{a}^\dagger\right] &=& \hat{a}^\dagger
\end{eqnarray}
where $\hat{n} = \hat{a}^\dagger\hat{a}$ is the number operator.\\
We introduce a new family of operators
\begin{equation}
    \tilde{\Pi}(\alpha,\beta) \equiv \exp(-i(\alpha\cdot\hat{a}^\dagger -
    \beta\cdot\hat{a}))
\end{equation}
Then
\begin{equation}   
    \tilde{\Pi}(\alpha,\beta)\tilde{\Pi}(\alpha',\beta') =
    \tilde{\Pi}(\alpha+\alpha',\beta+\beta')\tilde{Q}(\alpha,\beta;
    \alpha'\beta')
\end{equation}
where
\begin{equation}
    \tilde{Q}(\alpha,\beta;\alpha',\beta') = \exp(\frac{1}{2}(\alpha \beta'
    -\beta\alpha'))
\end{equation}
Thus we once again have the same structure as before -- not surprisingly, the
transformation $(q,p)\rightarrow (a^\dagger,a)$ is merely a rotation -- 
but
note the absence of the imaginary unit in $\tilde{Q}$, this is of course due to
the absence of an $i$ in the fundamental commutator relations in this basis.\\
The importance of this example is the following:
\begin{itemize}
    \item Fermions can be described by a similar algebra, but with
    anti-commutators; the quantities $\alpha,\beta$ then become Grassmann
    numbers. (This will be shown later.)
    \item We can treat fields by letting the operators carry a continous index
    (an element in some vector space or manifold) and inserting 
    delta-functions where appropriate.
    \item Any Lie algebra, finite or infinite dimensional, can be written in a
    form with creation and annihilation operators together with ``number
    operators'' (a root decomposition).
\end{itemize}
We should proceed with caution here. The algebra now consists of $3n+1$
generators, namely $\hat{a},\hat{a}^\dagger,\hat{n},1$, and while $1$ belongs
to the center, and thus can be ignored, this is by now means the case for
$\hat{n}$. Why not use
\begin{displaymath}
    \mbox{\P}(\alpha,\beta,\gamma) \equiv \exp(-i\alpha\cdot\hat{a}^\dagger+
      i\beta\cdot\hat{a}-i\gamma\cdot\hat{n})
\end{displaymath}
instead? This would clearly alter the relations:
\begin{eqnarray*}
    \mbox{\P}(\alpha,\beta,\gamma)\mbox{\P}(\alpha',\beta',\gamma') &=&
    \exp\left( -i(\alpha+\alpha')\cdot\hat{a}^\dagger+i(\beta+\beta')\cdot
      \hat{a}
    -i(\gamma+\gamma')\cdot\hat{n}-\right.\\
    &&\left. \frac{1}{2}(\alpha\cdot\beta'-\alpha'\cdot\beta)
    +(\alpha\cdot\gamma'-\alpha'\cdot\gamma)\hat{a}^\dagger-
    (\beta\cdot\gamma' - \beta'\cdot\gamma)
    \hat{a}+...\right)
\end{eqnarray*}
We note one thing: To any order the term involving the extra generator 
$\hat{n}$ looks like $i(\gamma+\gamma')\cdot\hat{n}$, there are no higher order 
terms. Nor does
it alter the symplectic product. The new generator only modifies the 
expression for the deformed addition, i.e. the terms involving $\hat{a},
\hat{a}^\dagger$. The $\gamma,\gamma'$ appears more or less as some arbitrary
parameters. The problem can be traced back to the fact that $\hat{n}$ is {\em 
not} and independent quantity. Dependent quantities will be
elements of the universal enveloping algebras, i.e. polynomials in the
generators, and should thus not be included among the basic quantities -- they
should be non-linear functions of the classical phase-space variables, and not
independent coordinates. This distinction will become clearer as we consider
semisimple Lie algebras in the sequel.

\subsubsection{Some Comments: Quantum Planes and Fibres}
We ellaborate a little bit on the structure involved in the WWM formalism as
outlined above. The essential quantity was seen to be the operator $\Pi(u,v)$.
This then lead to a deformation of the classical Poisson structure and to an
isomorphism between the Hilbert-Schmidt operators and the functions on
phasespace. Now, this deformation can also come about in another way. Define
\begin{equation}
    X = e^{\hat{q}} \qquad Y = e^{\hat{p}}
\end{equation}
then
\begin{equation}
    XY =qYX
\end{equation}
where $q=\exp(i\hbar)$. Hence $X,Y$ makes up a non-commutative geometry, known
as the {\em quantum plane} ${\bf {\sf R}}^2_q$ \cite{WZ}, which is a deformation of the 
classical space
${\bf {\sf R}}^2$. The automorphism group of this quantum plane is then what is known as a
{\em quantum group}, a deformed version of a classical Lie group.\\
Define now
\begin{equation}
    X(u) = e^{u\hat{q}} \qquad Y(u) = e^{u\hat{p}}
\end{equation}
then we have what we could call a {\em quantum fibre bundle} where the base
space is ${\bf {\sf R}}$ and the fiber at $u$ is a copy of the quantum plane. The
deformation parameter $q$ develops a $u$-dependency, so we have different
deformations at different points (the fibres are of course still isomorphic,
though). We further note the non-local ``folding''
\begin{equation}
    X(u)Y(v) = q(u,v)Y(v)X(u)
\end{equation}
which holds even when $u\neq v$. Let us finally note that $\Pi(u,v)$ is
essentialy just $X(v)Y(u)$. These arguments then indicate that quantum groups
will indeed appear upon quantization of classical theories. In fact, the entire
formalism as presented here is very intimately related to the study of quantum
groups, see e.g. \cite{EK} for a related study of deformation of Poisson-Lie 
algebras.

\section{An Arbitrary Lie Algebra}
We now want to generalize the WWM approach to the case where the given quantum
phase-space is an arbitrary Lie algebra. Two special cases are particularly
important, namely abelian and semisimple algebras, and will be treated first. 
Then we will comment on how to generalize to non-abelian, non-semisimple Lie
algebras.

\subsection{Abelian Lie Algebras}
For each natural number $n$ there exists just one (up to isomorphism) abelian
Lie algebra $\bf a$ with $\dim~{\bf a}=n$. And this Lie algebra is isomorphic 
to
${\bf {\sf F}}^n$, where ${\bf {\sf F}}$ is the base-field (e.g. the reals or the complex numbers).
The universal enveloping algebra $U({\bf a})$ can then be identified with
the ring of formal power series in $n$ (commuting) variables:
\begin{equation}
    U({\bf a}) = {\bf {\sf F}}\left[\left[X_1,...,X_n\right]\right]
\end{equation}
Thus we simply take the vector space ${\bf {\sf F}}^n$ to be our classical phase space
$\Gamma^0_{\bf a}$
\begin{equation}
    \Gamma^0_{\bf a} \equiv {\bf {\sf F}}^n \simeq {\bf a}
\end{equation}
Note, however, that the name ``phase-space'' is somewhat inappropriate as
$\Gamma^0_{\bf a}$ will in general not be a symplectic space - in fact it will
only be so if $n$ is even, in which case we have the canonical symplectic form
\begin{equation}
    \omega_0(X,X') \equiv X\wedge X' \equiv \sum_{i=1}^{n/2} 
    \left(X_iX_{i'+n/2} - X_{i+n/2}X_i'\right)
\end{equation}
All the same, for simplicity we will stick to the name phase-space even in the
case where $n=\dim~{\bf a}$ is odd.\\
We should notice that $\Gamma^0_{\bf a}$ is a flat manifold (it is a vector
space). It will turn out that non-abelian Lie algebras have non-flat 
phase-spaces. In the abelian case $C(\Gamma^0_{\bf a})$ is simply the space 
of all 
functions which have a formal Taylor expansion. In general, this will of 
course not be true.\\
As $\bf a$ is abelian so is $U({\bf a})$ and hence so is $C(\Gamma^0_{\bf a})$,
i.e. the twisted product is just the usual product of functions
\begin{equation}
    f(X)*g(X) = f(X)g(X)
\end{equation}
There is an analogy with the case of abelian $C^*$-algebras here: the famous
Gel'fand theorem \cite{BR,C*} states that any abelian $C^*$-algebra is 
isomorphic to either
the space $C_0(X)$ of continous functions vanishing at infinity or the space
$C_b(X)$ of bounded functions on some locally compact Hausdorff space $X$. We
will later come across suggestions that this relationship between the 
WWM-formalism for Lie algebras
as proposed here and the Gel'fand theory for $C^*$-algebras goes deeper
than this.\\
We can collect the above in the following
\begin{prop}
Let $\bf a$ be an abelian Lie algebra with $n=\dim~{\bf a} <\infty$ over some
field ${\bf {\sf F}}$, then
\begin{enumerate}
    \item the classical phase-space becomes $\Gamma_{\bf a}^0 \equiv {\bf {\sf F}}^n \simeq
    {\bf a}$, when $n$ is even this is a symplectic space,
    \item $C(\Gamma_{\bf a}^0)\simeq {\bf {\sf F}}[[X_1,...,X_n]]$ is the set of all formal
    power series in $n$ variables, and
    \item the twisted product on $C(\Gamma_{\bf a}^0)$ becomes trivial $f*g=fg$
\end{enumerate}
\end{prop}

\subsection{Semisimple Lie algebras}
Many models in physics use not only the Heisenberg algebra but also some finite
or infinite dimensional Lie algebra $\bf g$. The obvious examples are 
Yang-Mills theories, $\sigma$-models, current algebras, conformal field theory,
and string theory. In a Yang-Mills theory the fields $A_\mu$ (and their 
conjugate momenta $\pi^\mu$) are elements of some Lie algebra $\bf g$; $A_\mu =
A_\mu^k\lambda_k$ where $[\lambda_k,\lambda_l]=ic_{kl}^{~~m}\lambda_m$. The 
same
goes for $\sigma$-models, in current algebras we have commutator relations
between the various components of the currents, $[J_\mu^k(x), J_{\bf {\sf N}}u^l(x')]
=i\delta(x-x')\eta_{\mu{\bf {\sf N}}u}c^{kl}_{~~m}J_\mu^m(x)$. In conformal field theory we
have a family of fields $\phi_i(z,\bar{z})$ depending on two complex variables 
and satisfying the so-called {\em conformal bootstrap} \cite{Fuchs}
\begin{displaymath}
    \phi_i(z,\bar{z})\phi_j(w,\bar{w}) = d_{ij}^{~~k}(z,\bar{z},w,\bar{w})
    \phi_k(w,\bar{w})
\end{displaymath}
A similar situation arises in string theory. As we can see, this is more or less
the generic situation in modern physics, and hence we need to extend our WWM
formalism to phase-spaces extending the Heisenberg algebra.\\
For clarity we will first develop the formalism for finite dimensional
semi-simple Lie algebras, and then we will make the (rather straightforward)
generalization to their loop algebras and (affine) Kac-Moody algebras.\\
From basic Lie algebra theory (see e.g. \cite{Fuchs,Jacobson}) we know that 
we can choose a 
convenient basis $\{E_\alpha, H^i\}$ for the semisimple Lie algebra $\bf g$ 
such that
\begin{eqnarray}
    \left[H^i,H^j\right] &=& 0\\
    \left[H^i,E_\alpha\right] &=& \alpha^iE_\alpha\\
    \left[E_\alpha,E_\beta\right] &=& \left\{\begin{array}{ll}
    N_{\alpha,\beta}E_{\alpha+\beta} & \alpha+\beta\mbox{ a non-zero root}\\
    \alpha_iH^i & \alpha+\beta=0\\
    0 & \mbox{otherwise} \end{array}\right.    
\end{eqnarray}
where $N_{\alpha,\beta}$ are some constants. The elements $H^i, i=1,...,l$ 
span the Cartan subalgebra $\bf h$ of $\bf g$ and act as number operators. 
The remaining
elements $E_\alpha$ act as creation and annihilation operators (depending on 
the
sign of the root $\alpha$). When $\alpha$ is a root, so is $-\alpha$, hence we
can divide the elements $E_\alpha$ into pairs $E_{\pm\alpha}$.
We thus suggest the following generalization ($\alpha$ positive):
\begin{equation}
    a_i \mapsto E_{-\alpha} \qquad\qquad a_i^\dagger \mapsto E_{+\alpha}
    \qquad\qquad n_i \mapsto H^i
\end{equation}
As our basic translation operator $\Pi(u,v)$ ($u,v$ are now $r$-dimensional
vectors, where $\dim {\bf g} = n = 2r+l, l={\rm rank}~{\bf g} = \dim~{\bf h}$) 
we will thus use
\begin{deff}
If $\bf g$ is a semisimple Lie algebra of finite dimension and ${\bf g} = {\bf
g}_0+\sum_{\alpha>0}({\bf g}_\alpha+{\bf g}_{-\alpha})$ is a root decomposition
with respect to the Cartan subalgebra ${\bf g}_0$ then we define the Weyl map in
terms of
\begin{displaymath}
    \Pi(u,v) = \exp(iu^\alpha E_{+\alpha}-iv^\alpha E_{-\alpha}+i\lambda^j(u,v)
    H_j)
\end{displaymath}
summing over positive roots.
\end{deff}
In general we 
cannot {\em a priori} omit the Cartan element (it would in general not give
rise to a bijective map), so we have to include them explicitly, but, 
on the other
hand, they are the analogues of the number operators and should thus not be
counted as ``independent'' quantities, i.e. the parameters $\lambda^i$ should
not be independent coordinates but instead $\lambda^i=\lambda^i(u,v)$. These
dependent coordinates $\lambda^i$ are related to an imbedding of the phase-space
which is $n-l=2r$ dimensional into a $n$ dimensional vector space.\footnote{
We should also be aware of the fact that using the
matrix trace is perhaps not the most general procedure, instead one could 
define
an abstract trace as a linear functional $\chi$ with the property $\chi(AB) =
\chi(BA)$, as this implies $\chi([A,B])=0$ we see that the number of such
posible generalizations can be labeled by elements of the first cohomology 
class
$H^1({\bf g})$ of the Lie algebra $\bf g$. There will in general be 
essentially two, namely $\chi(AB)={\rm Tr}~A {\rm Tr}~B$ and $\chi(AB)={\rm Tr}~(AB)$.
The first of these must be discarded as it would imply $A_W\propto {\rm Tr}~A$
for all $A$, which is clearly unsatisfactory, hence only the second alternative
is usable.}\\
We cannot, however, simply take over the relation
\begin{displaymath}
    \Pi(u,v)\Pi(u',v') = \Pi(u+u',v+v')Q(u,v;u',v')
\end{displaymath}
instead it will turn out that the vector sum $u+u', v+v'$ gets deformed, as
does the symplectic product $\xi\wedge\xi' = uv'-vu'$.
Hence we can write ($\xi\equiv(u,v)$)
\begin{equation}
    \Pi(\xi)\Pi(\xi') = \Pi(\xi\oplus\xi')Q(\xi\times\xi')
\end{equation}
Here $Q$ depends only upon central and Cartan elements (for $\bf g$ semisimple,
and only upon elements in the maximal abelian subalgebra otherwise, as will be
explained later).\\
The extra non-commutativity of the phase-space leads to a deformation of the
vector-space structure of ${\bf {\sf R}}^{2r}$ the deformed vector sum being $\oplus$. 
The explicit form for $\xi\oplus \xi'$ is
found by using the Baker-Campbell-Hausdorff formula, but for simplicity we will
wait untill the example ${\bf g}=su_2$ below before we will write it out
explicitly. Note that this deformation of the vector space structure on
${\bf {\sf R}}^{2r}$ implies that the classical phase-space ($(u,v)$-space) might not be a
vector space, but just a manifold. We will denote it by $\Gamma$ or 
$\Gamma_{\bf g}$ when we wish to emphasize which algebra it belongs to. The 
symplectic product $\wedge$ gets deformed to $\times$.\\
The corresponding twisted product can be written in terms of a kernel $\Delta$
like
\begin{equation}
    (f*g)(\xi) = \int_\Gamma \Delta(\xi,\xi',\xi'') f(\xi')g(\xi'') d\xi'd\xi''
\end{equation}
where
\begin{eqnarray}
    \Delta(\xi,\xi',\xi'') &=& {\rm Tr}\left(\Pi(u,v)\Pi(u',v')\Pi(u'',v'')
    \right)\nonumber\\
    &=& {\rm Tr}\left(e^{i\langle\xi\oplus\xi'\oplus\xi'',
    E\rangle}e^{i(\xi\times(\xi'\oplus\xi'')+\xi'\times\xi'',H)}
    \right)
\end{eqnarray}
where we have defined
\begin{eqnarray*}
    \langle\xi,E\rangle &=& u^\alpha E_{+\alpha}-v^\alpha E_{-\alpha}\\
    (x,H) &=& x_jH^j
\end{eqnarray*}
In order to satisfy the same relations as for the Heisenberg algebra, we
must demand that ${\rm Tr}(\Pi(\xi)\Pi(\xi')) \equiv K(\xi,\xi')$ is a
reproducing kernel for $L^2(\Gamma)$. This is seen by inserting the definitions
of $A_W, B_W$ in
\begin{displaymath}
    \int_\Gamma A_W(\xi)B_W(\xi)d\xi = {\rm Tr}(AB)
\end{displaymath}
which allow us to express expectation values in terms of integrals over the
classical phase-space (let for instance $B=\rho$, the density matrix).\\
We have proven the following
\begin{prop}
Let $\bf g$ be as in Definition 1 above, then
\begin{enumerate}
    \item $\dim\Gamma = \dim {\bf g}-\dim{\bf g}_0 = n-l$
    \item writing $\xi=(u,v)$ we have $\Pi(\xi)\Pi(\xi') = \Pi(\xi\oplus \xi')
    Q(\xi\times \xi')$ with $Q$ only involving the Cartan elements
    \item the deformed addition is given by
    \begin{displaymath}
    \left(\begin{array}{c}u\\v\\ \lambda(u,v)\end{array}\right)\oplus
    \left(\begin{array}{c}u'\\v'\\ \lambda(u',v')\end{array}\right) = 
    \left(\begin{array}{c}u+u'+\mbox{higher order terms}\\ v+v'+\mbox{higher
    order terms}\\ \lambda(u,v)+\lambda(u',v')\end{array}\right)
    \end{displaymath}
    whereas the deformed symplectic product is
    \begin{displaymath}
    \xi\times\xi' = \omega_0(u,v,u',v')+\mbox{higher order terms}
    \end{displaymath}
    with 
    \begin{displaymath}
    \omega(u,v,u',v') \equiv \sum_{\alpha>0}(u_\alpha v'_\alpha-u'_\alpha
    v_\alpha)
    \end{displaymath}
\end{enumerate}
\end{prop}
Concerning the nature of $C(\Gamma)$ and products of $\Pi$ with itself we can 
say
\begin{prop}
Let $\Pi$ be the ``translation'' operator defining the Weyl map, then the
twisted product of two functions $f,g\in C(\Gamma)$ can be written in term of a
kernel $\Delta$
\begin{displaymath}
    (f*g)(\xi) = \int f(\xi')g(\xi'')\Delta(\xi,\xi',\xi'') d\xi' d\xi''
\end{displaymath}
where $d\xi$ is a measure invariant under the action of the corresponding Lie
group. The kernel is given by
\begin{displaymath}
    \Delta(\xi,\xi',\xi'') = {\rm Tr}~\Pi(\xi)\Pi(\xi')\Pi(\xi'')
\end{displaymath}
Furthermore, $K(\xi,\xi')$ given by
\begin{displaymath}
    K(\xi,\xi') = {\rm Tr}~\Pi(\xi)\Pi(\xi')
\end{displaymath}
is a reproducing kernel for $L^2(\Gamma)$.
\end{prop}

Before continuing with Kac-Moody algebras, let me comment on the suggested
formalism and its relations with other authors' proposals. Several authors have
studied the natural symplectic structure associated with a Lie algebra, see for
instance \cite{Liesymp}, this symplectic structure is based on the 
{\em coadjoint orbit action}. Given a Lie group $G$, we construct the 
symplectic space
$O_m = \{m'={\rm Ad}^*(g)m~|~ g\in G\}$, where $m$ is some point. The 
symplectic structure is given by the {\em Kirilov-Kostant Poisson bracket}
\begin{displaymath}
    \{f,g\}_{\em KKP}(m) \equiv \langle m,\left[df(m),dh(m)\right]\rangle
\end{displaymath}
here $\langle\cdot,\cdot\rangle$ denotes the pairing between ${\bf g}$, the 
Lie algebra of $G$, and its dual ${\bf g}^*$. Kasperkowitz \cite{Kasper} 
have applied this symplectic
structure to the WWM formalism. These proposals are relevant when the 
``coordinate manifold'' is a Lie algebra and one then needs to find a
phase-space. For an arbitrary coordinate manifold $M$ (i.e. $q$-space) the
associated phase-space is the cotangent bundle $T^*M$, so even if the global
momentum space ($p$-space) is not defined, the phase-space {\em is} 
well-defined. It
is this construction the coadjoint orbit formalism generalizes for $M$ replaced
by an arbitrary Lie algebra. But, {\em a priori}, systems do exist for which 
the phase-space
cannot be rewritten as a cotangent bundle, i.e. phase-spaces do exist for which
we can define neither a global coordinate manifold nor a global momentum space.
Darboux' theorem \cite{geomq} asserts, though, that we can always define 
coordinates $p,q$
locally satisfying the usual Poisson bracket relations. The generalization of
the WWM formalism proposed here, is able to handle this situation easily as it
is based directly on the phase-space manifold and not on the coordinate
manifold. What we, in this paper, are essentially doing is to reconstruct a 
topological space $\Gamma$ by a ring of continous functions $C(\Gamma)$ on 
it (i.e. essentially ``point-less topology'', or perhaps rather ``point-less 
differential geometry'').\\
Before the example, which will hopefully clarify the formalism somewhat, let me
just briefly mention infinite dimensional Lie algebras. Given a finite
dimensional Lie algebra $\bf g$ with generators $\lambda^k$, we can construct
the corresponding infinite dimensional Lie algebra of maps $S^1\rightarrow
{\bf g}$, this algebra is known as the {\em loop algebra} of $\bf g$, and 
will be
denoted by ${\bf g}_{\rm loop}$. A basis for this Lie algebra is $\lambda^k_m =
\lambda^kz^m$ where $z$ is a complex number of modulus 1. The commutator
relations are
\begin{equation}
    \left[\lambda^k_m,\lambda^l_n\right] = ic^{kl}_{~~h}\lambda^h_{m+n}
\end{equation}
This is probably the simplest way of generating infinite dimensional Lie
algebras. The more general class of {\em Kac-Moody algebras} \cite{Fuchs,Kac} 
is based on a relaxation of the restraints on the Cartan matrix 
$A^{ij}$, interestingly this too leads to infinite dimensional Lie algebras. 
An important subclass of these algebras, the so-called {\em affine Kac-Moody 
algebras} (defined by demanding the Cartan matrix to be positive semi-definite)
can be viewed as a non-trivial central extension of a loop algebra, and a 
basis can be chosen such that
\begin{eqnarray}
    \left[ H^i_m,H^j_n\right] &=& mG^{ij}\delta_{m+n,0}K\\
    \left[ H^i_m,E^\alpha_n\right] &=& \alpha^i E_{m+n}^\alpha\\
    \left[ E_m^\alpha,E_n^\beta\right] &=& N_{\alpha\beta}
    E^{\alpha+\beta}_{m+n}\\
    \left[ E^\alpha_m,E^{-\alpha}_{-m}\right] &=& \alpha_i H^i_m+mK
\end{eqnarray}
where $\alpha,\beta$ are roots, $N_{\alpha\beta}=0$ if $\alpha+\beta$ is not a
root, $G^{ij}$ is some matrix and $K$ is the central generator. The eigenvalue
of $K$ is known as the {\em level}. Notice that the generators with $m=n=0$ 
span
a subalgebra, which is an ordinary Lie algebra. Affine Kac-Moody algebras 
can be
included in our formalism by making the substitution $u_i\mapsto u_i^m, i=1,
2,...,r; m=0,\pm 1, \pm 2,...$ so each $u_i$ gets replaced by an entire
sequence leading to an infinite dimensional classical phase-space. In order to
deal with non-affine Kac-Moody algebras, we will have to go back to the general
commutator relations, as no particular representation in terms of other 
algebras
are known. If we just treat $A^{ij}$ as an arbitrary matrix we can include also
these kinds of Kac-Moody algebras in our formalism -- in principle at least.

\section{An Example: $su_2=so_3$}
To really see the formalism at work, we will consider the simplest non-trivial
example, namely ${\bf g}=su_2$. For simplicity we will work in the $s=1/2$ 
representation only
(later we will show that the result is independent of the choice of
representation),
the generators can then be chosen to be the Pauli matrices $\sigma_i$, from
which we can define
\begin{displaymath}
    \sigma_\pm = \frac{1}{\sqrt{2}}(\sigma_1\pm i\sigma_2)
\end{displaymath}
But it will be just as easy to work directly with $\sigma_i$ instead and we 
will do this.\\
The ``translation'' operator is then
\begin{equation}
    \Pi(u,v) = \exp(iu\sigma_1-iv\sigma_2+i\lambda(u,v)\sigma_3)
\end{equation}
which can be rewritten as (using the familiar properties of the Pauli matrices)
\begin{equation}
    \Pi(u,v) = \cos\sqrt{u^2+v^2+\lambda^2}+i(u\sigma_1-v\sigma_2+\lambda
      \sigma_3)
    \frac{\sin\sqrt{u^2+v^2+\lambda^2}}{\sqrt{u^2+v^2+\lambda^2}}
\end{equation}
The most important ingredient is the deformed addition and symplectic product.
Defining $\xi=(u,v)$ and
\begin{equation}
    \xi\wedge\xi' \equiv uv'-vu'
\end{equation}
the usual $h_n$-case would read
\begin{displaymath}
    \Pi(\xi)\Pi(\xi') = \Pi(\xi+\xi')Q(\xi\wedge\xi')
\end{displaymath}
with
\begin{displaymath}
    Q(\xi\wedge\xi')=e^{i\xi\wedge\xi'}
\end{displaymath}
This gets deformed to
\begin{equation}
    \Pi(\xi)\Pi(\xi') = \Pi(\xi\oplus\xi')Q(\xi\times\xi')
\end{equation}
where $\oplus$ is the deformed vector sum and $\times$ the deformed symplectic
product 
\begin{eqnarray}
    \xi\oplus\xi' &=& \xi+\xi'+\mbox{cubic terms}\\
    \xi\times\xi' &=& \xi\wedge\xi'+\mbox{quartic terms}
\end{eqnarray}
Computing the first corrections we get
\begin{equation}
    \xi\oplus\xi' = \xi+\xi'+\frac{1}{3}(\xi\wedge\xi')(\xi'-\xi) +
    \mbox{higher order terms}
\end{equation}
Now, it follows from the properties of the Pauli matrices 
\begin{displaymath}
    e^{i\sigma_j u} = \cos u+i\sigma_j\sin u
\end{displaymath}
that the function $\Pi$ can be expressed in terms of trigonometric functions
so we must demand periodicity in the arguments. This implies that the classical
phase space, $\Gamma$, can be one of two spaces (upto diffeomorphism), 
namely the torus $S^1\times S^1$ or the
sphere $S^2$. It is the commutator relations which determines which of the two
spaces we have. Our phase-space cannot be written as a product space $U\times
V$, where $u\in U, v\in V$, as $[\sigma_+,\sigma_-]=2\sigma_3 \not\in 
Z({\bf g})$ ($Z({\bf g})$ denotes the center of the Lie algebra) and hence the
classical phase-space must be $S^2$, as we would expect \cite{su2}. 
The torus would correspond to a Lie algebra
\begin{eqnarray*}
    \left[E_+,E_-\right] &=& 0\\
    \left[H,E_+\right] &=& aE_+\\
    \left[H,E_-\right] &=& -bE_-
\end{eqnarray*}
where $a,b$ are arbitrary positive numbers. A more rigorous argument is given in
the section on general properties.\\
The requirement ${\rm Tr}(AB) = \int_\Gamma A_W B_Wd\xi$ together with ${\rm
Tr}(A)<\infty$ for all $A$ in the universal enveloping algebra of $su_2$,
implies that $\|A_W\|_2^2= \int_\Gamma |A_W|^2d\xi<\infty$ for all $A_W\in
C(\Gamma)$. Thus $C(\Gamma)\simeq L^2(S^2)$.\\
This shows that, although
the classical phase-space inherits an addition making it locally isomorphic to
the vectorspace ${\bf {\sf R}}^{2r}$, this isomorphism will in general only be local. Thus
{\em the classical phase-space will be some $2r$-dimensional real, symplectic
manifold}. The global topological structure of this manifold could 
(a priori) be
representation dependent -- we will return to this point later -- but the 
example 
suggests that only the commutator relations matter. The essential point 
is\footnote{This actually only holds with some slight modifications as will
be explained later.}
\begin{displaymath}
    \mbox{non-commutativity } \longrightarrow \mbox{ non-flatness}
\end{displaymath}
We can write the ``translation'' operator $\Pi$ as
\begin{equation}
    \Pi(u,v) = f_0(u,v) + \sigma\cdot f(u,v)
\end{equation}
with
\begin{eqnarray*}
    f_0(u,v) &=& \cos \sqrt{u^2+v^2+\lambda^2}\\
    f_1(u,v) &=&iu\frac{\sin\sqrt{u^2+v^2+\lambda^2}}
    {\sqrt{u^2+v^2+\lambda^2}}\\
    f_2(u,v) &=&-iv\frac{\sin\sqrt{u^2+v^2+\lambda^2}}
    {\sqrt{u^2+v^2+\lambda^2}}\\
    f_3(u,v) &=&i\lambda\frac{\sin\sqrt{u^2+v^2+\lambda^2}}
    {\sqrt{u^2+v^2+\lambda^2}}\\
\end{eqnarray*}
The Weyl maps of the generators become
\begin{eqnarray}
    (1)_W &=& 2f_0(u,v)\\
    (\sigma_i)_W &=& 2f_i(u,v)
\end{eqnarray}
The factors of two can be removed by multiplying the trace by $1/(2s+1)$.
We must demand $f_0\equiv const$, which is the same as requiring
$u^2+v^2+\lambda^2=const$, i.e. we once again get $\Gamma\simeq S^2$.
Normalizing such that $(1)_W=1$ we get
\begin{equation}
    u^2+v^2+\lambda^2=\arccos^2\frac{1}{2}
\end{equation}
which, then gives $\lambda$ as a function of $u,v$.\\
We notice that, had we taken $\lambda=0$ we would have arrived at the 
most unfortunate result $(\sigma_3)_W=0$, i.e. we would map the non-abelian 
algebra $su_2$ onto an abelian one. Instead we have $\lambda=\pm \sqrt{const^2
-u^2-v^2}\neq 0$. We note that to lowest order the generators $\sigma_1,
\sigma_2$ (or equivalently
$\sigma_\pm$) gets mapped to $u,v$, whereas $(\sigma_3)_W$ is quadratic, to
lowest order, in $(u,v)$. This is because the Cartan subalgebra of a semisimple
Lie algebra can be obtained from the root spaces ${\bf g}_{\pm \alpha}= {\bf {\sf F}}
E_{\pm \alpha}$ as $\left[{\bf g}_\alpha,{\bf g}_{-\alpha}\right] \subseteq 
{\bf g}^0={\bf h}$. The Cartan elements are in this way not truely independent
quantities.\\
The reproducing kernel $K(u,v;u',v')$ and the kernel of the twisted product
$\Delta$ becomes
\begin{eqnarray}
    \frac{1}{2}K&=& 1-f_j(u,v)f_k(u',v')\delta^{jk}\\
    \frac{1}{2}\Delta &=& 1-f_j(u,v)f_k(u',v')\delta^{jk}-f_j(u,v)f_k(u'',v'')
    \delta^{jk}-f_j(u',v')f_k(u'',v'')\delta^{jk}+\nonumber\\
    &&\sum_{ijk}f_i(u,v)f_j(u',v')f_k(u'',v'')
\end{eqnarray}
The proposed WWM-formalism has a very beautiful representation in terms of
wellknown quantities. For the sake of generality we will work in a general
irreducible representation corresponding to an angular momentum $l$. The
translation operator can be expanded
\begin{equation}
    \Pi(u,v) = \sum_{mm'} \Pi_{mm'}(u,v)|lm\rangle\langle lm'|
\end{equation}
where
\begin{equation}
    \Pi_{mm'}(u,v) \equiv \langle lm'|\Pi(u,v)|lm\rangle = \langle lm' |
    e^{iu\sigma_1-iv\sigma_2+i\lambda\sigma_3}|lm\rangle 
    \equiv D_{m'm}^l(R_{(u,v)})
\end{equation}
where $R_{(u,v)}$ is the rotation given by the angles $u,v$. The $D_{mm'}^l(R)$
is the usual representation matrices for rotations \cite{Merz}. For ${\bf
g}=h_n$, the Heisenberg algebra, $\Pi(\xi)$ constituted a (ray-)
representation of the group of translations, whereas for ${\bf g}=su_2$ we 
get a (proper)
representation of the group of rotations, the phase-space became the orbits 
of these groups, i.e. the plane and the sphere respectively.\\ 
The Weyl map of an ``operator'' (i.e. a $(2l+1)\times(2l+1)$-matrix) $A$ becomes
\begin{equation}
    A_W(u,v) = \sum_{mm'}D_{m'm}^l(R_{(u,v)})\langle lm|A|lm'\rangle \equiv
    \sum_{mm'}D^l_{m'm}A_{mm'}
\end{equation}
A very beautiful result. At this point we should notice that our WWM-formalism
is slightly different from the ``standard approach'' developed by V\'{a}rilly
and Gracia-Bond\'{i}a, \cite{su2}. Our formulas are slighty simpler, as we do 
not have Clebsch-Gordon coefficients occuring explicitly. Their
``translation''-operator, which they denote by $\Delta^l$, is essentially our
translation operator $\Pi$, in fact $\Delta^{1/2} \sim Y_{00}+\Pi$.\\
The inverse Weyl-map of a function is also interesting to compute. Let 
$f(u,v) = \sum_m f_m(u,v)|lm\rangle$ be some function in $C(S^2)$, then 
the corresponding operator, which we will denote by $f^W$ is simply
\begin{equation}
    f^W = \sum_{mm'}\int D_{m'm}^l(R_{(u,v)})f_m(u,v) d\Omega |lm\rangle 
    \langle lm'|
\end{equation}
where $d\Omega$ denotes the measure on $S^2$.\\
We can also use the rotation matrices $D_{mm'}^l$ to write
\begin{equation}
    \Delta(\xi,\xi',\xi'') = {\rm Tr}~\Pi(\xi)\Pi(\xi')\Pi(\xi'') = \sum_m
    D_{mm}^l(R_\xi R_{\xi'}R_{\xi''})
\end{equation}
Let us finish this subsection by making a comment on the measure on $\Gamma$.
Clearly this measure $d\mu$ has to satisfy a few requirements: (1) it must be a
Borel measure (the $\sigma$-algebra must be given by the topology, such that
continous functions are measurable), (2) it must be a Radon measure, i.e. the
measure of a bounded set is bounded, and finally (3) it must be invariant under
the group $G$ (i.e. Haar), which the operators $\Pi(\xi)$ constitute a 
representation of. For
the Heisenberg algebra this implies that $d\mu$ is the usual Lebesgue-measure,
as this is the only translation invariant Radon measure (upto a multiplicative
constant), whereas for $su_2$ it implies that $d\mu=d\Omega$, the usual solid
angle measure.

\subsection{The Corresponding Loop and Kac-Moody Algebras}
Let us also consider the corresponding loop algebra $(su_2)_{\rm loop}$, which
will be our first example of an infinite dimensional Lie algebra. The 
commutator relations are
\begin{equation}
    \left[\sigma_j^n,\sigma_k^m\right] = 2i\varepsilon_{jk}^{~~l}\sigma_l^{n+m}
\end{equation}
where $\sigma_j^n=\sigma_j z^n$ with $z\in S^1$. Let $\bar{\sigma}_j$ denote 
the sequence $\{\sigma_j^n\}_{n\in{\bf {\sf Z}}}$ and define $\bar{u}=\{u_n\}_{n\in{\bf {\sf Z}}}$. 
We then introduce our, by now familiar, translation operator 
\begin{equation}
    \Pi_{\rm loop}(\bar{u},\bar{v}) \equiv \exp\left(i\left(\bar{u}\cdot
    \bar{\sigma}_+-\bar{v}\cdot\bar{\sigma}_-+\bar{\lambda}\cdot\bar{\sigma}_3
    \right)\right)
\end{equation}
where
\begin{equation}
    \sigma_\pm^n \equiv \frac{1}{\sqrt{2}}\left(\sigma_1^n\pm i\sigma_2^n
    \right)
\end{equation}
with the obvious notation
\begin{displaymath}
    \bar{u}\cdot\bar{\sigma_j} \equiv \sum_{n=-\infty}^\infty u_n\sigma_j^n
\end{displaymath}
In terms of the basis $\{\sigma_\pm^n,\sigma_3^m\}$ the commutator relations 
are
\begin{displaymath}
    \left[ \sigma_+^n,\sigma_-^m\right] = 2\sigma_3^{n+m} \qquad \qquad \left[
    \sigma_3^n,\sigma_\pm^m\right] = \pm \sigma_\pm^{n+m}
\end{displaymath}
and we have
\begin{equation}
    \Pi_{\rm loop}(\xi)\Pi_{\rm loop}(\xi') = \Pi_{\rm loop}(\xi\oplus\xi')
    Q_{\rm loop}(\xi\times\xi')
\end{equation}
Now
\begin{equation}
    \bar{u}\cdot\bar{\sigma}_j \equiv \sum_{n=-\infty}^\infty u_n\sigma_j^n =
    \left(\sum_{n=-\infty}^\infty u_nz^n\right)\sigma_j \equiv u(z)\sigma_j
\end{equation}
so the translation operator for the loop algebra can be expressed in terms of
that of the basic Lie algebra as
\begin{equation}
    \Pi_{\rm loop}(\xi) = \Pi(\xi(z))
\end{equation}
where $\xi(z) = \sum_n \xi_nz^n$ is an analytic function $S^1\rightarrow\Gamma=
S^2$. This is a general result. The classical phase-space of the loop algebra 
is
the space of functions $S^1\rightarrow \Gamma$, where $\Gamma$ is the classical
phase-space belonging to the original Lie algebra. Symbolically
\begin{equation}
    \Gamma({\bf g}_{\rm loop}) \equiv 
    \Gamma(C^\infty(S^1\rightarrow{\bf g})) \simeq  
    C^\infty(S^1\rightarrow \Gamma({\bf g}))
\end{equation}
The deformation function $Q_{\rm loop}$ can be expressed in terms of $Q$ as
\begin{equation}
    Q_{\rm loop}(\xi\times\xi') = Q(\xi(z)\times\xi'(z))
\end{equation}
where
\begin{equation}
    \xi(z)\times\xi'(z) \equiv \sum_{n,m=-\infty}^\infty (u_nv'_m - v_nu_m')
    z^{n+m}
\end{equation}
Thus the generalization to the loop algebra of a given Lie algebra is trivial.\\
The Kac-Moody algebra $\widehat{su_2}$ at level $k$ can be obtained from the
loop algebra as
\begin{eqnarray*}
    \left[\sigma_3^n,\sigma_3^m\right] &=& km\delta_{n,-m}\\
    \left[\sigma_3^n,\sigma_\pm^m\right] &=& \pm \sigma_\pm^{n+m}\\
    \left[\sigma_+^n,\sigma_-^m\right] &=& 2\sigma_3^{n+m}+km
\end{eqnarray*}
The translation operator is defined to be
\begin{equation}
    \Pi_{\rm KM}(\xi) = \Pi_{\rm loop}(\xi) \label{eq:KM}
\end{equation}
but with this new non-trivial central extension it satisfies
\begin{equation}
    \Pi_{\rm KM}(\xi)\Pi_{\rm KM}(\xi') = \Pi_{\rm KM}(\xi\oplus\xi') Q_{\rm KM}
    (\xi\times\xi')
\end{equation}
The deformation function $Q_{\rm KM}$ differs from $Q$ by terms proportional to
$k$, its $\sigma_3^n$ term is identical to that of
the loop algebra, which means that $Q_{\rm KM}$ differs from $Q$ by a C-number
function:
\begin{equation}
    Q_{\rm KM}(\xi\times\xi') = {\cal Q}_k(\xi,\xi') Q(\xi(z)\times\xi'(z))
\end{equation}
Explicitly
\begin{equation}
    {\cal Q}_k(\xi,\xi') = 1-k\sum_{n=-\infty}^\infty n(u_nv_n'-u_n'v_n) +O(k^2)
\end{equation}
This is also a general result; for an arbitrary Lie algebra ${\bf g}$ each
element $u_n,v_n$ would be $r$-dimensional, $u_n=(u_n^1,...,u_n^r$) etc., and 
we have to include a sum over this extra index in the above formula too, but
otherwise the analysis holds.\\
We have now seen how the proposed formalism works for a simple example, ${\bf
g}= su_2$. Furthermore, we have seen how to relate the WWM formalism for a loop
algebra or a Kac-Moody algebra to that of the original algebra, by which these
infinite dimensional algebras are generated.\\
As a final comment we should note that the relationship (\ref{eq:KM}) 
implies that the
two classical phase-spaces, $\Gamma_{\rm loop}, \Gamma_{\rm KM}$, will be 
identical, the correspondence rules (the Weyl
maps) will be different though, and, in the language of an earlier subsection,
so would their corresponding quantum fibre bundles. We can summarize this in the
following
\begin{prop}
Let $\bf g$ be a finite dimensional semisimple Lie algebra, and denote by ${\bf
g}_{\rm loop}$ and $\hat{{\bf g}}_k$ its corresponding loop and affine Kac-Moody
algebra at level $k$ respectively. The corresponding classical phase-spaces are
denoted by $\Gamma_{\bf g}, \Gamma({\bf g}_{\rm loop})$ and $\Gamma(\hat{{\bf
g}}_k)$ respectively and their ``translation'' operators by $\Pi, \Pi_{\rm
loop}$ and $\Pi_{\rm KM}$, then
\begin{enumerate}
    \item $\Gamma({\bf g}_{\rm loop}) \simeq C^\infty (S^1\rightarrow
    \Gamma_{\bf g})$
    \item $\Pi_{\rm loop}(\xi) = \Pi(\xi(z))$ and $Q_{\rm loop}(\xi\times\xi') =
    Q(\xi(z)\times\xi'(z))$ with $z\in S^1$ and
    \begin{displaymath}
        \xi(z)\times \xi'(z) = \sum_{n,m=-\infty}^\infty (u_nv'_m-u'_nv_m)
        z^{n+m}+\mbox{higher order terms}
    \end{displaymath}
    \item $\Gamma(\hat{{\bf g}}_k)\simeq \Gamma({\bf g}_{\rm loop})$
    \item $\Pi_{\rm KM}(\xi) = \Pi_{\rm loop}(\xi)$ and $Q_{\rm KM}(\xi\times
    \xi') = {\cal Q}_k(\xi,\xi')Q_{\rm loop}(\xi\times\xi')$ where ${\cal Q}_k$
    depends on the level $k$ as
    \begin{displaymath}
    {\cal Q}_k(\xi,\xi') = 1-k\sum_{n=-\infty}^\infty n(u_nv'_n -u'_n v_n)+
    O(k^2)
    \end{displaymath}
\end{enumerate}
\end{prop}
There is an immediate generalization of the loop algebras to the gauging of
any finite dimensional Lie algebra. The algebra of local gauge transformations
is locally\footnote{The group is not given by this simple formula globaly, 
since we do not take the principal bundle structure into account; globally, 
the correct group is the group preserving the corresponding principal bundle,
see e.g. \cite{Nash}. For simplicity, though, we will consider only this 
particular group, $C^\infty(M)\otimes{\bf g}$, also sometimes denoted by
${\rm Map}(M,{\bf g})$.}
\begin{equation}
    \tilde{\bf g}(M) = C^\infty(M\rightarrow {\bf g}) = C^\infty(M)\otimes 
{\bf g}
\end{equation}
wherefrom
\begin{equation}
    \Pi_{\tilde{\bf g}(M)}(u,v) = \Pi_{\bf g}(u(x),v(x))\qquad x\in M
\end{equation}
and we have the following
\begin{corr}
With $\bf g$ a semisimple Lie algebra of finite dimension and $M$ any manifold
we have
\begin{equation}
    \Gamma(C^\infty(M)\otimes{\bf g}) \simeq C^\infty(M\rightarrow\Gamma) =
    C^\infty(M)\otimes \Gamma({\bf g})
\end{equation}
Borrowing a word from the theory of $C^*$-algebras we could call $C^\infty(M)
\otimes {\bf g}$ the {\em suspension} of $\bf g$. The result above then reads:
{\em The phase-space of a suspension is the suspension of the phase-space.}
\end{corr}

\section{The structure of the Classical Phase-Space}
Now, the classical phase-space was constructed from a map $\Pi(u,v)$, and
clearly it is closely related to the Lie groups with $\bf g$ as their Lie
algebra. In fact, had $\lambda$ been independent of $(u,v)$ we would have 
gotten a local Lie group \cite{LieI}. Let $G$ be the smallest connected 
Lie group with $\bf g$
as its Lie algebra (note, that $G$ might not be simply-connected), this then
acts transitively on $\Gamma$, and thus, \cite{LieI,LieIII}, 
$\Gamma\simeq G/H_0$, where $H_0$ is
some subgroup. Hence the classical phase-space is a homogenous space. From the
construction it follows that $H_0$ is essentially a Lie group with $\bf h$, the
Cartan subalgebra, as its Lie algebra, it is not, however, identical to simply
$\exp({\bf h})$ as we have to subtract the center. Hence $H_0=H\backslash Z$
where $H$ is the smallest connected Lie group with $\bf h$ as its Lie algebra.
Very often we have only a trivial center, so often $H_0=H$.
For ${\bf g}=su_2=so_3$ we thus have $G=SO_3$ and $H=SO_2$, whereby (trivial
center)
\begin{displaymath}
    \Gamma_{su_2} \simeq\Gamma_{so_3}\simeq SO_3/SO_2\simeq SU_2/U_1 \simeq S^2
\end{displaymath}
as we saw earlier.\\
We notice that for $\bf g$ semisimple, $\bf h$, and thus also $H$, will be
abelian, whereas for a more general Lie algebra it will just be nilpotent. We
can consider $H$ as the subgroup spanned by the diagonal matrices, when $G$ 
is a
matrix group. The case of semisimple Lie algebras simplifies enormously by the
abelianness of the Cartan group, since any abelian Lie group has the form 
${\bf {\sf F}}^n
\times {\bf {\sf T}}^m$, where ${\bf {\sf F}}$ is the base field and $\bf {\sf 
T}$ is the torus (${\bf {\sf T}}=S^1$, 
i.e. essentially $SO_2$ or $U_1$). Hence for compact Lie groups $H={\bf {\sf 
T}}^l$.\\
We should furthermore notice that a homogenous space is symplectic if it is of
the form $G/H_\omega$ where $H_\omega$ is the connected component of the kernel
of some antisymmetric two-form $\omega$, \cite{GeomI}. An obvious such 2-form is
\begin{equation}
    \omega_0(u,v,\lambda,u',v',\lambda') = \left(\begin{array}
      {c}u\\v\end{array}\right)
    \wedge \left(\begin{array}{c}u'\\v'\end{array}\right)
\end{equation}
where $\wedge$ is the canonical symplectic product on ${\bf {\sf R}}^{2r}$. Clearly $H=
{\rm
Ker}~\omega_0$. As we have seen, $\omega_0$ gets deformed to another
antisymmetric 2-form $\omega$, which can be found order by order from the
Baker-Campbell-Hausdorff theorem. This new 2-form
will again vanish on $H$ and nowhere else, and hence $\Gamma$ is indeed a
symplectic manifold when $\bf g$ is semisimple. Thus
\begin{prop}
    For $\bf g$ a semisimple Lie algebra with $n=\dim{\bf g}<\infty$ with Cartan
    subalgebra ${\bf g}_0$ we have
    $\Gamma\simeq G/H$ where $G, H$ are the smallest, connected Lie groups 
    with ${\bf g}, {\bf g}_0$ as their Lie algebras. Furthermore, $\Gamma$ is
    symplectic.
\end{prop}
Now, this was based on the assumption that $\bf g$ was semisimple. 
For an arbitrary Lie algebra, this will not be
the case. In general the Cartan subalgebra is defined as a maximal
nilpotent subalgebra which is its own normalizer, i.e.
\begin{eqnarray}
    \underbrace{\left[{\bf g},\left[{\bf g},...,\left[{\bf g},{\bf g}\right]
    ...\right]\right]}_{n \mbox{ brackets}} &=& 0 
    \mbox{  (for a sufficiently large $n$)}\\
    \{ x\in {\bf g}~|~\left[x,{\bf h}\right]\subseteq {\bf h}\} &=& {\bf h}
\end{eqnarray}
For any representation $\rho:{\bf g}\rightarrow gl(V)$, where $V$ is some 
vector space, we can then write \cite{LieI,LieIII}
\begin{equation}
    V = \oplus_{i=1}^r V^{\lambda_i}
\end{equation}
where
\begin{equation}
    V^{\lambda} = \{v\in V~|~ \exists m\in{\bf {\sf N}} \forall x: (\rho(x)-\lambda(x))^m
    v=0\}
\end{equation}
The quantities $\lambda$ are linearly independent functionals on $\bf h$, i.e.
$\Phi_\rho = \{\lambda_1,...,\lambda_r\} \subseteq{\bf h}^*$; they are the 
weights. A root is then defined as a non-zero weight in the adjoint
representation, i.e. $\Delta = \Phi_{\rm ad}\backslash \{0\}$. We still have a
root decomposition
\begin{equation}
    {\bf g} = {\bf h}\oplus\left(\oplus_{\alpha\in\Delta}{\bf g}_\alpha \right)
\end{equation}
and
\begin{eqnarray}
    \left[{\bf g}_\alpha, {\bf g}_\beta\right] && \left\{\begin{array}{ll}
    \subseteq {\bf g}_{\alpha+\beta} & \alpha+\beta\in\Phi\\
    =0 & \alpha+\beta\not\in\Phi\end{array}\right.\\
    B({\bf g}_\alpha,{\bf g}_\beta) &=& 0\qquad \alpha+\beta\neq 0
\end{eqnarray}
where $B(\cdot,\cdot)$ is the Killing form. Hence we still have some degree of
orthogonality of the different root spaces. Unfortunately it no longer holds
that $\alpha\in\Delta\Rightarrow-\alpha\in\Delta$, so the roots nolonger come in
pairs. Thus the classical phase-space, which we can still define as we {\em do}
have a root decomposition, will nolonger be even-dimensional, and {\em a
fortiori} not symplectic, in the general case. Hence
\begin{displaymath}
    {\bf g}\mbox{ semisimple}\qquad\Rightarrow\qquad \Gamma_{\bf g}\mbox{ symplectic}
\end{displaymath}
The use of the Cartan algebra as suggested above would constitute one
generalization to non-semisimple algebras, but I would like to propose 
another one, which I think is more appropriate. The reason for the succes of 
the formalism in the
semisimple case can be traced back to the fact that for such algebras the
maximal nilpotent and the maximal abelian subalgebra coincide: that the Cartan
algebra becomes abelian. So it was actually the abeliannes of $\bf h$ that was
used. Furthermore, while Cartan algebras of semisimple Lie algebras are fairly
unique (they are conjugate) this will not in general hold for 
Cartan subalgebras
of general Lie algebras, whereas abelian Lie algebras are characterized
completely by the dimension and thus are unique (up to isomorphism). 
So what I propose to do is
consider not a maximal nilpotent Lie subalgebra $\bf h$, but a maximal abelian
subalgebra $\bf a$. Now, clearly abelian Lie algebras are also nilpotent so we
can use the above decomposition (which actually only holds for complex Lie
algebras and not in general for real ones) for any (real or complex or
otherwise) Lie algebra $\bf g$. The dimensionality $s=\dim~{\bf a}$ will not,
however, be equal to the rank $l$ of the Lie algebra. Let us call this 
number for the {\em abelian rank}, written ${\rm a-rank}~({\bf g})$. Obviously
\begin{equation}
    1+\dim~ Z({\bf g})\leq {\rm a-rank}~({\bf g}) \leq {\rm rank}~({\bf g})
\end{equation}
Let $\Phi$ denote the set of weights $\lambda_i$ in the adjoint representation,
and let $\Delta=\Phi\backslash\{0\}$, then we once more have a decomposition
\begin{equation}
    {\bf g} ={\bf g}_0\oplus\left(\oplus_{\alpha\in\Delta}{\bf g}_\alpha\right)
\end{equation}
with ${\bf a}={\bf g}_0$ and
\begin{eqnarray}
    \left[{\bf g}_\alpha,{\bf g}_\beta\right] &&\left\{\begin{array}{ll}
    =0 & \alpha+\beta\not\in\Phi\\
    \subseteq {\bf g}_{\alpha+\beta} & \alpha+\beta\in\Phi
    \end{array}\right.\\
    \left[{\bf g}_0,{\bf g}_\alpha\right] &\subseteq & {\bf g}_\alpha
\end{eqnarray}
We should notice that this construction implies that two Lie algebras have the
same classical phase space if and only if one is the central extension of the
other or the one can be written as the direct sum of the other and an abelian
algebra. In other words abelian algebras get mapped to the singleton set
$\{0\}$. This of course differs from the definition of $\Gamma^0$ for an abelian
algebra given earlier, but agrees with our calculations for $su_2$. In fact,
this is the reason why we inserted the superscript $0$ in the definition of the
abelian case. Furthermore, this implies that our formalism assigns the same
classical phase-space (upto isolated points, which can always be discared on
physical grounds) to two algebras ${\bf g}_1,{\bf g}_2$ ($\dim{\bf g}_2
\geq\dim{\bf g}_1$, say) which differ by the addition of an abelian algebra
$\bf a$ (i.e. ${\bf g}_2={\bf g}_1+{\bf a}$) such that $[{\bf a},{\bf g}_1]
\subseteq Z({\bf g}_2)$, for instance when ${\bf g}_2$ is a central extension 
of ${\bf g}_1$ or when the sum is direct. The only exception to this is when 
${\bf g}_1$, say, is itself abelian, then $\Gamma_{{\bf g}_2}\simeq
\Gamma^0_{{\bf g}_1}$,
so the formalism is consistent with our choice of phase-space for an abelian 
Lie algebra -- an example is of course the Heisenberg algebra, which is a 
central extension of ${\bf {\sf R}}^{2n}$. Note, however, that even though the
classical phase-space coincide, their correspondence rules given by the
operators $\Pi_{1,2},Q_{1,2}$ differ as will their quantum fibre bundles.
\begin{prop}
Two finite dimensional Lie algebras ${\bf g}_1, {\bf g}_2$ have the same
classical phase-spaces upto isolated points if and only if one is the semidirect
sum of an abelian algebra $\bf a$ and the other, say ${\bf g}_2={\bf g}_1+{\bf
a}$, with $[{\bf a},{\bf g}_1]\subseteq Z({\bf g}_2)$. A special case is when
${\bf g}_2$ is a central extension of ${\bf g}_1$.
\end{prop}

\subsection{Nilpotent and Solvable Lie Algebras}
Some particular important cases of non-semisimple Lie algebras are the 
nilpotent
and solvable algebras. Let us make a few comments on the WWM formalism of 
these. Recall that a Lie algebra, $\bf g$, is solvable if its derived series,
$({\bf g}^{(i)})$, with ${\bf g}^{(i)} = \left[{\bf g}^{(i-1)}, 
{\bf g}^{(i-1)}\right]$ for
$i \geq 1$ and ${\bf g}^{(0)}={\bf g}$, becomes trivial after a certain 
number of steps, i.e. ${\bf g}^{(i)}=0$ for some value of $i$. Similarly, a
Lie algebra is nilpotent if the series $({\bf g}_{(i)})$ with ${\bf g}_{(i)}
=\left[{\bf g},{\bf g}_{(i-1)}\right]$ becomes trivial after a certain number
of steps. A nilpotent Lie algebra is also solvable, and any Lie algebra can be
written as the semidirect sum of a solvable and a semisimple Lie algebra
(Levi-decomposition). Hence once we know how to deal with solvable algebras
we can in principle handle {\em any} Lie algebra.\\
As far as solvmanifolds (i.e. homogenous spaces of a solvable Lie
group) are concerned, let me just mention that both the M\"{o}bius band and the 
Klein bottle are both solvmanifolds, and that any solvmanifold can be 
written as a fibre bundle
over a compact solvmanifold with fibre ${\bf {\sf R}}^k$ for some $k$ (see \cite{LieI}). 
When the manifold
is even a nilmanifold (i.e. when $G$ is nilpotent), then this fibre bundle
can be trivialized. Indeed, if $M$ is any nilmanifold, then \cite{LieI}
\begin{equation}
	M\simeq M^*\times {\bf {\sf R}}^n
\end{equation}
where $M^*$ is a compact nilmanifold. If $M=G/H$, then $M^*=~^aH/H$, where 
$^aH$ denote the {\em algebraic closure} (i.e. the closure in the Zariski
topology) of $H$. Hence, when $H$ comes from the maximal abelian subalgebra
of $\bf g$, the Lie algebra of $G$, then $^a{\bf h} = {\bf h}$, so $^aH/H$ is
discrete, i.e.
\begin{equation}
	\Gamma \simeq {\bf {\sf R}}^n\times\mbox{discrete group}
        \qquad n=\dim {\bf g}-\dim{\bf h}
\end{equation}
This makes the case of nilpotent Lie algebras very simple (as we already 
noticed when we dealt with the Heisenberg algebra).\\
One should notice, that we can obtain solvable Lie algebras from nilpotent
ones by the following exact sequence
\begin{equation}
	0\rightarrow {\bf g}'\rightarrow {\bf g}\rightarrow {\bf g}/{\bf g'}
	\rightarrow 0
\end{equation}
When $\bf g$ is solvable, then ${\bf g'}$ is the nil-radical, i.e. the largest
nilpotent subalgebra. Thus solvable Lie algebras can be gotten as extensions
of nilpotent Lie algebras by abelian ones. We will return to extensions when
we deal with $C^*$-algebras.\\
Now, {\em a priori} the suggested WWM-map will not be a bijection for
non-semisimple Lie algebras, as we do not {\em a priori} have ${\bf g}_0
\subseteq \cup_{\alpha,\beta\in\Delta}\left[{\bf g}_\alpha,{\bf g}_\beta
\right]$. Algebras for which this does happen will be referred to as {\em good}
algebras, whereas the rest will be termed {\em defficient}. The defficiency can
be labeled by an integer $\delta({\bf g};{\bf g}_0) = 
\dim\{x\in{\bf g}_0~|~\forall
\alpha,\beta\in\Delta~:~x\not\in\left[{\bf g}_\alpha,{\bf g}_\beta\right]\}$.
For semisimple Lie algebras we have $\delta=0$. For defficient algebras it can
happen that for some representations the proposed WWM-map is bijective whereas
for others it is not. The task of classifying good/defficient algebras and
``good'' representations for defficient ones will not be undertaken here; my
main interest lies with semisimple algebras. For the remaining of this paper,
then, only good algebras will be considered. A priori, different copies of the
maximal abelian subalgebra could have different defficiencies. It is always
understood that the one which minimizes $\delta({\bf g};{\bf g}_0)$ is to be
chosen, i.e. we chose the one with the maximal overlap with the derived algebra.
This is summarized in the following
\begin{deff}
Let $\bf g$ be a finite dimensional Lie algebra and let ${\bf g}_0$ be the
maximal abelian subalgebra (unique upto ismorphisms). The defficiency is
\begin{displaymath}
    \delta({\bf g};{\bf g}_0) = \dim\{x\in {\bf g}_0~|~\forall \alpha,\beta \in
    \Delta~:~x\not\in [{\bf g}_\alpha,{\bf g}_\beta]\}
\end{displaymath}
where $\delta$ denotes the set of roots in a decomposition w.r.t. ${\bf g}_0$.
It is always understood that the copy of ${\bf g}_0$ which minimizes $\delta$ is
to be used. With this, writing $\Delta=\Delta_+\cup\Delta_-$ where $\Delta_+$
consists of positive roots, $\Delta_-$ of negative roots, the ``translation''
operator becomes
\begin{displaymath}
    \Pi(u,v) = \exp(i\sum_{\alpha\in\Delta_+}u_\alpha E_\alpha-i\sum_{\alpha \in
    \Delta_-}v_\alpha E_\alpha+i\lambda^j(u,v)H_j)
\end{displaymath}
where $H_j$ generate ${\bf g}_0$.
\end{deff}

\section{Some Further Examples}
We saw that the classical phase-space of $su_2=so_3$ turned out to be $S^2$. 
Let us now consider a few more examples very briefly.\\
Let us start with the Lie algebra of the non-compact group $SU(1,1)$, it
consists of traceless $2\times 2$ matrices (in the fundamental 
representation) which obey
\begin{equation} 
    XJ=-JX^\dagger\qquad\qquad J=\left(\begin{array}{cc}1 & 0\\ 0 & -1
    \end{array}\right) = \sigma_3
\end{equation}
The commutator relations are
\begin{eqnarray}
    \left[ H,X_1\right] &=& -2X_2\\
    \left[ H,X_2\right] &=& -2X_1\\
    \left[X_1,X_2\right]&=& -2iH
\end{eqnarray}
And a representation is 
\begin{eqnarray*}
    X_1 &=& \left(\begin{array}{cc} 0 & 1 \\ 1 & 0 \end{array}\right)\\
    X_2 &=& \left(\begin{array}{cc} 0 & i \\ -i & 0\end{array}\right) =
    -\sigma_2\\
    H &=& iJ = i\sigma_3
\end{eqnarray*}
We can get from a representation of $su_2$ to one of $su_{1,1}$ by making the
transformation (a ``Wick rotation'')
\begin{equation}
    \sigma_1 \mapsto \sigma_1=X_1 \qquad \sigma_2\mapsto -\sigma_2 = X_2 \qquad
    \sigma_3\mapsto i\sigma_3 = H
\end{equation}
Inserting this in $\Pi(u,v)$ we get
\begin{equation}
    \Pi_{su_{1,1}}(u,v) = e^{iu\sigma_1+iv\sigma_2-\lambda \sigma_3}
\end{equation}
For the $su_2$-case we could introduce spherical coordinates for $(u,v,\lambda)
$, here it turns out that we get the following coordinates
\begin{eqnarray*}
    u &=& z\cos\alpha \cosh\beta\\
    v &=& z\sin\alpha \cosh\beta\\
    \lambda &=& z\sinh\beta
\end{eqnarray*}
allowing us to write
\begin{equation}
    \Pi_{su_{1,1}}(u,v) = \cos z+i(\cos\alpha\cosh\beta \sigma_1 +\sin\alpha
    \cosh\beta\sigma_2+i\sinh\beta \sigma_3)\sin z
\end{equation}
And the classical phase space becomes
\begin{equation}
    \Gamma(su_{1,1}) \simeq \left\{(u,v,\lambda)\in {\bf {\sf R}}^3~|~ u^2+v^2-\lambda^2 =
    const. \right\} \equiv S^{1,1}
\end{equation}
i.e. a hyperboloid.\\
Now, from $su_2$ and $su_{1,1}$ we can construct a number of important Lie
algebras, by noting \cite{LieIII} $so_4=su_2\oplus su_2, so_{2,2} = 
su_{1,1}\oplus 
su_{1,1}$ and $u_2^*({\bf {\sf H}}) = su_2\oplus su_{1,1}$ where ${\bf {\sf H}}$ denote the ring of
quarternions. The Lie algebra $so_{3,1}$, the Lorentz algebra, can also be
constructed by noting $so_{3,1}=sl_2({\bf {\sf C}})_{\bf {\sf R}} = su_2\oplus i\cdot su_2 = 
su_2\otimes{\bf {\sf C}}$, where
$sl_2({\bf {\sf C}})_{\bf {\sf R}}$ means $sl_2({\bf {\sf C}})$ considered as a real algebra. These Lie algebras
consists of $4\times 4$ matrices of the form
\begin{eqnarray*}
    so_4\simeq su_2\oplus su_2 &\qquad& \left(\begin{array}{cccc}
        0 & \alpha & \beta & \gamma\\
        -\alpha & 0 & a & b\\
        -\beta & -a & 0 & c\\
        -\gamma & -b & -c & 0 \end{array}\right) \qquad (a,b,c), (\alpha,\beta,
        \gamma) \in {\bf {\sf R}}^3\\
    so_{3,1}\simeq sl_2({\bf {\sf C}})_{\bf {\sf R}} &\qquad &\left(\begin{array}{cccc}
        0 & i\alpha & i\beta & i\gamma\\
        -i\alpha & 0 & a & b\\
        -i\beta & -a & 0 & c\\
        -i\gamma & -b & -c & 0 \end{array}\right)\qquad (a,b,c), (\alpha,\beta,
        \gamma) \in {\bf {\sf R}}^3\\
    so_{2,2} \simeq su_{1,1}\oplus su_{1,1} &\qquad&\left(\begin{array}{cccc}
        0 & x & i\alpha & i\beta\\
        -x & 0 & i\gamma & i\delta\\
        -i\alpha & i\gamma & 0 & z\\
        -i\beta & -i\delta & -z & 0 \end{array}\right) \qquad x,z \in {\bf {\sf R}},
        \alpha,\beta,\gamma,\delta \in {\bf {\sf R}}\\
    u_2^*({\bf {\sf H}}) \simeq su_2\oplus su_{1,1} &\qquad& \left(\begin{array}{cccc}
        0 & x & a & b\\
        -x & 0 & \bar{b} & d\\
        -a & -\bar{b} & 0 & \bar{x}\\
        -b & -d & -\bar{x} & 0 \end{array}\right) \qquad x,b\in {\bf {\sf C}}, a,d\in {\bf {\sf R}}
\end{eqnarray*}
We must thus find
an expression for $\Gamma_{{\bf g}_1\oplus{\bf g}_2}$. Let us start with
$so_4=su_2\oplus su_2$. We simply get
\begin{eqnarray}
    \Pi_{so_4}(u_1,v_1,u_2,v_2) &=& \Pi_{su_2}(u_1,v_1)\Pi_{su_2}(u_2,v_2)\\
    Q_{so_4}(u_1,v_1,u_2,v_2) &=& Q_{su_2}(u_1,v_1)Q_{su_2}(u_2,v_2)
\end{eqnarray}
This is a general result:
\begin{prop}
If ${\bf g}_1, {\bf g}_2$ denote two Lie algebras then
\begin{eqnarray}
    \Pi_{{\bf g}_1\oplus{\bf g}_2} &=& \Pi_{{\bf g}_1}\Pi_{{\bf g}_2}\\
    Q_{{\bf g}_1\oplus{\bf g}_2} &=& Q_{{\bf g}_1}Q_{{\bf g}_2}
\end{eqnarray}
Similarly, if $\bf g$ can be written as the sum of two Lie algebras
with $\left[{\bf g}_1,{\bf g}_2\right] \in Z({\bf g})$  then
\begin{eqnarray*}
    \Pi_{\bf g} &=& \Pi_{{\bf g}_1}\Pi_{{\bf g}_2}\\
    Q_{\bf g} &=& Q_{{\bf g}_1}Q_{{\bf g}_2}q_Z
\end{eqnarray*}
where $q_Z$ some element in $\exp(Z({\bf g}))$. It also follows from this that
\begin{equation}
    \Pi_{\bf g} =\Pi_{{\bf g}/{\bf h}} \Pi_{\bf h}
\end{equation}
when $\bf h$ is any ideal in $\bf g$. Thus the classical phase-spaces become
\begin{eqnarray}
    \Gamma_{{\bf g}_1\oplus{\bf g}_2} &=& \Gamma_{{\bf g}_1}\times\Gamma_{{\bf
    g}_2}\\
    \Gamma_{{\bf g}_1+{\bf g}_2} &=& \Gamma_{{\bf g}_1}\times \Gamma_{{\bf
    g}_2} \mbox{   (when }\left[{\bf g}_1,{\bf g}_2\right] \in Z({\bf g})
      ~{\rm )}
\end{eqnarray}
\end{prop}    
We should emphasize once more that the
classical phase-spaces of an algebra and its central extensions are isomorphic
(upto isolated points),
the correspondence between algebra and functions on phase-space is different,
though, and hence so are the corresponding quantum fibre bundles. Such central
extensions are of great importance when ${\bf g}_1= {\bf g}_2$, the algebra 
$\bf g$ is then a {\em Heisenberg double} of ${\bf g}_1$.\footnote{In fact, 
for ${\bf g}_1={\bf g}_2={\bf {\sf R}}$ we get the usual Heisenberg algebra. This shows 
that the new correspondence which the central extension introduces, can be 
seen as related to quantization.} In a typical gauge theory,
for instance, we have two set of operators $\phi_k,\pi_k$ which both of them
span some Lie algebra ${\bf g}_1$ at each point $x$ and each instant $t$. The
algebra is not just the gauging of ${\bf g}_1\oplus{\bf g}_1$, but a central
extension of it as we have to impose $[\phi_k(x,t),\pi_j(x',t')]_{t=t'} =
i\delta(x-x')\delta_{jk}$, the canonical relation.\\
For the algebras just mentioned we have at once
\begin{eqnarray}
    \Gamma_{so_4} &=& S^2\times S^2\\
    \Gamma_{so_{2,2}} &=& S^{1,1}\times S^{1,1}\\
    \Gamma_{u_2^*({\bf {\sf H}})} &=& S^2\times S^{1,1}
\end{eqnarray}
The Lorentz algebra is somewhat more complicated. It arrises as a
complexification of $su_2$, and there is thus a non-trivial automorphism
exchanging the real and complex parts of a Lie element. This means that
\begin{equation}
    \Gamma_{so_{3,1}} = \frac{SO_3\times SO_3}{SO_2\times SO_2}
\end{equation}
where $SO_2\times SO_2$ is imbedded in some non-trivial way in $SO_3\times 
SO_3$
because of this automorphism. But noting that $so_{3,1}$ is thus a 
complexification of $su_2$, i.e. $so_{3,1}=su_2\otimes{\bf {\sf C}}$, we get
\begin{equation}
	\Gamma_{so_{3,1}}=\Gamma_{su_2\otimes{\bf {\sf C}}} \simeq \Gamma_{su_2}\otimes{\bf {\sf C}} 
= S^2\otimes{\bf {\sf C}}
\end{equation}
i.e. we can view the phase-space of a complexification as a kind of 
``complexification'' of the original phase-space.\\
Let us now move on to a Lie algebra of rank two, namely $su_3$, represented 
by the Gell-Mann matrices
$\lambda_i, i=1,..,8$. We would expect the classical phase space to have a
dimensionality of $8-2=6$. The key ingredient in the $su_2$ case was the useful
relation $\sigma_i\sigma_j = i\epsilon_{ij}^{~~k}\sigma_k$, which allowed us to
get a nice expression for $\Pi(u,v)$ in terms of trigonometric functions. For
$su_3$ we can use
\begin{eqnarray}
    \left[\lambda_a, \lambda_b \right] &=& if_{ab}^{~~c}\lambda_c\\
    \left\{\lambda_a, \lambda_b\right\}&=& \frac{4}{3}\delta_{ab} +
    2d_{ab}^{~~c}\lambda_c
\end{eqnarray}
where $f_{abc}$ is totally antisymmetric, whereas $d_{abc}$ is totally
symmetric. From this it follows that
\begin{equation}
    \lambda_a\lambda_b = if_{ab}^{~~c}\lambda_c+\frac{2}{3}\delta_{ab} +
    d_{ab}^{~~c}\lambda_c \label{eq:su3prod}
\end{equation}
Thus any function $f$ of the generators can be written as
\begin{equation}
    f(\lambda) = f_0 + \lambda_a f^a
\end{equation}
where $f_0,f^a$ are complex numbers, independent of the generators. These can 
be obtained from $f$ by taking traces:
\begin{eqnarray*}
    f_0 &=& \frac{1}{3}{\rm Tr}~f(\lambda)\\
    f^a &=& \frac{1}{3}{\rm Tr}(f(\lambda)\lambda^a)
\end{eqnarray*}
Particularly useful for us are monomials $(u\cdot\lambda)^n$, we write
\begin{equation}
    (u\cdot\lambda)^n = a_n(u)+\lambda_a b_n^a(u)
\end{equation}
the coefficients satisfying
\begin{eqnarray}
    a_{n+1} &=& \frac{2}{3} u\cdot b_n\\
    b_{n+1}^a &=& a_n u^a+u^bb_n^cd^a_{~~bc}
\end{eqnarray}
with $a_0=1, a_1=0, b_0^a=0, b_1^a=u^a$. Explicitly, the kernel $\Delta$ and the
translation operator $\Pi$ becomes
\begin{eqnarray}
    \Pi(u) &=& c_0(u)+\lambda^ac_a(u)\\
    \Delta(u,v,w) &=& c_0(u)c_0(v)c_0(w)+\frac{2}{3}\delta^{ab}\sum_{\rm perm}
    c_a(u)c_b(v)c_0(w)+\nonumber\\
    &&\frac{2}{3}(d_{abc}+if_{abc})c^a(u)c^b(v)c^c(w)
\end{eqnarray}
where
\begin{eqnarray*}
    c_0(u) &\equiv & \sum_{n=0}^\infty \frac{i^n}{n!}a_n(u)\\
    c^a(u) &\equiv & \sum_{n=0}^\infty \frac{i^n}{n!}b_n^a(u)
\end{eqnarray*}
The product of two translation operators becomes
\begin{eqnarray}
    \Pi(u)\Pi(v) &=& c_0(u)c_0(v)+\frac{2}{3}\delta^{ab}c_a(u)c_b(v)+\nonumber\\
    &&\lambda^c\left(c_0(u)c_c(v)+c_c(u)c_0(v)+(if^{ab}_{~~c}+d^{ab}_{~~c}) 
      c_a(u)c_b(v)\right)
\end{eqnarray}
whereby the reproducing kernel, in this representation, reads
\begin{equation}
      K(u,v) = c_0(u)c_0(v)+\frac{2}{3}\delta^{ab}c_a(u)c_b(v)
\end{equation}
The classical phase-space becomes
\begin{equation}
    \Gamma_{su_3} = SU_3/S(U_1\times U_1\times U_1) = SU_3/U_1\times U_1
\end{equation}
In general
\begin{equation}
      \Gamma_{su_n} = SU_n/S(U_1^n) = SU_n/U_1^{n-1}
\end{equation}
with $U_1^k=U_1\times...\times U_1$ ($k$ factors). I do not think these 
homogenous spaces have any name.\\
We can get some insight into the structure of $\Gamma_{su_3}$ by evaluating the
Weyl symbols of the generators. Using (\ref{eq:su3prod}) and the fact that the
generators are traceless, one easily sees (the factor of two can of course be
removed by a suitable normalization of the trace)
\begin{eqnarray}
    (1)_w &=& 2c_0(u)\\
    (\lambda_a)_W &=& 2c_a(u)\\
\end{eqnarray}
thus we must oncve more demand $c_0=const$, which imposes a constraint on the 
variable $u^a$, deforming the phase-space from
simply ${\bf {\sf R}}^6$ to some 6-manifold, just like for $su_2$ where the requirement
$f_0=const$ implied $\Gamma_{su_2}\simeq S^2$.\\
Furthermore, the symbol of a product becomes
\begin{equation}
    (\lambda_a\lambda_b)_W = 2c_0(u)\delta_{ab}+2(if_{ab}^{~~c}+d_{ab}^{~~c})
    c_c(u)
\end{equation}
comparing this with
\begin{equation}
    (c_a*c_b)(u) = \int c_a(v)c_b(w)\Delta(u,v,w)dvdw 
\end{equation}
we get
\begin{eqnarray}
    \delta_{ab} &=& \frac{3}{2}c_0^2\int c_a(v)dv\int c_b(w)dw+\nonumber\\
    &&\delta^{cd}\int c_c(v)c_a(v)dv~\int c_d(w)c_b(w)dw\\
    (d_{ab}^{~~c}+if_{ab}^{~~c})&=& \frac{1}{3}c_0
    \int(c_c(v)+c_c(w))c_a(v)c_b(w)dvdw+\nonumber\\
    &&\frac{1}{3}(d_{a'b'}^{~~~c}+if_{a'b'}^{~~~c})\int c^{b'}(v)c^{a'}(w)c_a(v)
    c_b(w)dvdww
\end{eqnarray}
which gives us some insight into the nature of the functions $c_a(u)$.\\
As an example of an infinite dimensional Lie algebra we can consider the 
{\em Witt algebra}, i.e. the algebra of diffeomorphisms of the circle. The 
commutator relations are
\begin{equation}
      \left[ A_n, A_m\right] = (m-n)A_{m+n}\qquad n,m\in{\bf {\sf Z}}
\end{equation}
Our largest abelian subalgebra is the one generated by $A_0$, hence
\begin{equation}
      \Pi(u,v) = \exp\left(i\sum_{n>0}(u_nA_n-v_nA_{-n})+i\lambda A_0\right)
\end{equation}
Now, from $A_n^\dagger = A_{-n}$ we see that $\bar{u}_n=v_n$, hence the 
classical phase-space consists of sequences $(u_n,v_n)$ of complex numbers 
such that $\bar{u}_n=v_n$ and $n=1,2,3...$. These can be represented just 
aswell by sequences $(x_n)$ with $x_n\in {\bf {\sf C}}, n\in{\bf {\sf Z}}$ satisfying $x_0=0$ and 
$\bar{x}_n=x_{-n}$, which again can be interpeted as a Fourier series, i.e.
\begin{eqnarray}
      \Gamma_{\rm Witt} &=& \{f\in C^\infty(S^1)~|~ f(\theta) = 
      \sum_{n=1}^\infty(c_n e^{in\theta}+c_{-n}e^{-in\theta}\}\\
      &=& \{f\in C^\infty(S^1)~|~f(0)=0\}
\end{eqnarray}
This contains also the spaces of $L^p$ functions on $S^1$ which vanish at 
$\theta=0$. The deformed sum is seen to be
\begin{eqnarray}
      &&\hspace{3cm}\left(\begin{array}{c}u_k\\ v_k\end{array}\right) \oplus 
      \left(\begin{array}{c}u'_k \\v'_k \end{array}\right) =\nonumber\\
      &&\hspace{-10mm}
      \left(\begin{array}{c} u_k+u'_k+\frac{1}{2}i\sum_{n=1}^{k-1}(k-2n)u_k
      u'_{k-n}-\frac{1}{2}ik(u_k\lambda'-u_k'\lambda)+
      \frac{1}{2}ik\sum_{n=1}^{k-1}(u_nv_{n-k}'-v_{n-k}u_n')+...\\
      v_k+v_k'+\frac{1}{2}i\sum_{n=1}^{k-1}(k-2n)v_kv'_{n-k}
      +\frac{1}{2}ik(v_k\lambda'-v_k'\lambda)-
      \frac{1}{2}ik\sum_{n=1}^{k-1}(v_nu_{n-k}'-u_{n-k}v_n')+...
      \end{array}\right)\nonumber\\
\end{eqnarray}
and the deformed symplectic product to be
\begin{equation}
      \left(\begin{array}{c}u\\v\end{array}\right)\times
      \left(\begin{array}{c}u'\\v'\end{array}\right) =
      \sum_k k(v_ku_k'-u_kv_k')+...
\end{equation}
similar to the result we found for the loop and Kac-Moody algebras.\\
Now, the {\em Virasoro algebra}
\begin{equation}
      \left[L_n,L_m\right] = (m-n)L_{n+m}+\delta_{n,-m} c_n
\end{equation}
is just a central extension of the Witt algebra and will hence have the same 
classical phase-space. We have seen earlier that also the classical 
phase-spaces of the loop algebras of semi-simple Lie algebras and their 
corresponding Kac-Moody algebras could be interpreted as function spaces over
the unit circle $S^1$. We will encounter more function spaces when we move on 
to consider $C^*$-algebras aswell.\\
Let us also briefly consider a defficient Lie algebra. The simplest algebra in
which ${\bf g}_0\cap{\bf g}'=\emptyset$ is the two-dimensional solvable Lie
algebra $[h,x]=x$, here the only weight is $\alpha=1$. A simple representation
is $h=x\frac{d}{dx}, x=x$. The dimensionality of the classical phase-space is
one, and from the non-commutativity we see that we can take $\Gamma\simeq 
S^1$. This algebra has been considered by Isham {\em et al.}, \cite{Isham},
in the context of developing a general quantization algorithm for non-trivial 
phase-spaces.\\
Some final important examples are the Poincar\'{e} algebra $iso(3,1)$ and the
Galilei algebra $gal_3$. The Poincar\'{e} algebra is the semidirect sum of
${\bf {\sf R}}^4$ and $so(3,1)=su_2\otimes{\bf {\sf C}}$. Clearly ${\bf {\sf R}}^4$ is the maximal abelian
subalgebra, and we get a classical phase-space of dimension $10-4=6$. In fact
the space must essentially be $SU_2\cdot SU_2\simeq S^3\cdot S^3$, where the 
dot
denotes some kind of product. It is rather surprising that the dimensionality
becomes six and not eight as one would have expected\footnote{This might be due
to the mass-shell constraint $p^2=m^2$ for the four-momentum together with the
 requirement that the particle move along a time-like geodesic, though.} and, 
furthermore, that it
is a kind of product of two compact manifolds. For the Galilei algebra we get
similarly a six dimensional phase-space (as it in this case was to be 
expected), but this time $SU_2\cdot{\bf {\sf R}}^3\simeq S^3\cdot {\bf {\sf R}}^3$, i.e. the limit 
$c\rightarrow
\infty$ which leads from the Galilei algebra from the Poincar\'{e} algebra
($c$ is the velocity of light), leads to an ``unwrapping'' of one $S^3$, or,
equivalently, that the finiteness of the velocity of light leads to a
compactification of ${\bf {\sf R}}^3$. This suggests that In\"{o}n\"{u}-Wigner 
contractions leads to a ``decompactification'' of the classical phase-space.\\
We have succeeded in obtaining Lie algebras yielding a number of two 
dimensional
manifolds as their classical phase-spaces as shown in table 1. We would like to
suggest that any surface can be obtained in this way, and as an example we will
construct a Lie algebra with the M\"{o}bius band as its classical phase-space.
The algebras in table 1 exhaust all non-trivial three dimensional Lie algebras,
hence the dimensionality of the wanted Lie algebra must be at leat four. Since
the M\"{o}bius band is a solvmanifold but not a nilmanifold, this algebra must
be solvable but not nilpotent. On the other hand, the cylinder and the
M\"{o}bius band differ only in the latter being a non-trivial bundle, but 
otherwise they both have the same local structure ${\bf {\sf R}}\times_{\rm loc}S^1$, 
where the subscript on
$\times_{\rm loc}$ is there to remind us that the product is only local in
general. So let us start with the algebra behind the cylinder
\begin{displaymath}
    \left[h,e\right] = e\qquad\left[h,f\right]=\left[e,f\right]=0
\end{displaymath}
and let us add a fourth generator $g$ mixing $e,f$,
\begin{displaymath}
    \left[g,e\right]=\alpha f\qquad\left[g,f\right]=\beta e
\end{displaymath}
The Jacobi identity then implies $\alpha=0$. We furthermore find ${\bf
g}'=\{e\}$, i.e. ${\bf g}''=0$ so the algebra is solvable, while ${\bf g}^n={\bf
g}'$ so the algebra is not nilpotent. The largest abelian subalgebra is ${\bf
h}={\rm span}~\{h,g\}$, and hence the dimensionality of the classical 
phase-space is
indeed two. Since $\Gamma$ is a solvmanifold of dimension two it has the form of
a (non-trivial) fibre bundle with fiber ${\bf {\sf R}}$ over some compact, one 
dimensional manifold $M_1$
\begin{displaymath}
    \Gamma \simeq {\bf {\sf R}}\times_{\rm loc} M_1
\end{displaymath}
and it is easy to see that the only possibility is $M_1=S^1$, wherefrom we get
\begin{equation}
    \Gamma \simeq \mbox{M\"{o}bius band}
\end{equation}
One could then go on to find Lie algebras corresponding to surfaces of genus
more than one, and, furthermore, to relate the topological characteristics
(Euler number, Stiefel-Whitney classes) to algebraic properties of the Lie
algebras -- a kind of generalized index theorem -- a point I plan to return to
in a sequel paper.

\section{Fermionic Degrees of Freedom}
Fermions are described by {\em anticommuting} creation and annihilation
operators
\begin{eqnarray}
    \left\{a_i, a_j\right\} &=& \left\{a_i^\dagger,a_j^\dagger\right\} =0\\
    \left\{a_i,a_j^\dagger\right\} &=& \delta_{ij}
\end{eqnarray}
We have no classical phase-space at our disposal. So we cannot construct an 
isomorphism between an algebra of operators and a Hilbert space of functions on
some vector-space (or manifold), i.e. as a space of functions with C-number
arguments. Rather, we have to define Grassmann numbers (which we will also 
refer to as G-numbers), abstract quantities satisfying
\begin{displaymath}
    \left\{\theta_i,\theta_j\right\} = \left\{\bar{\theta}_i,\bar{\theta}_j
    \right\} = \left\{\theta_i,\bar{\theta}_j\right\} =0
\end{displaymath}
We can treat these as ``coordinates'' and their corresponding differential
operators $\partial_i, \bar{\partial}_i$ as the ``momentum'' variables.\\
The generalization is now straightforward.
\begin{deff}
For fermionic creation- and annihilation-operators $a,a^\dagger$ we put
\begin{equation}
    \Pi(\theta,\eta) \equiv \exp(i\theta a^\dagger-i\eta a) 
\end{equation}
where $\theta,\eta$ are G-numbers anticommuting with the second quantization
operators aswell.
\end{deff}
This operator will the be our basis for
developing a WWM-formalism for fermionic degrees of freedom. The following
proposition is trivial:
\begin{prop}
The ``translation'' operator satisfies
\begin{equation}
    \Pi(\theta,\eta)\Pi(\theta',\eta')=\Pi(\theta+\theta',\eta+\eta')Q(\theta,
    \eta;\theta'\eta')
\end{equation}
where
\begin{equation}
    Q(\theta,\eta;\theta',\eta') =\exp(\theta\eta'+\eta\theta')
\end{equation}
\end{prop}
We notice that this is in fact a C-number, being the product of two G-numbers.
We also note that the sign in this G-symplectic product differs from the
symplectic product of two C-numbers. No deformation of the sum or the symplectic
product occurs here as the G-numbers are nilpotent $\theta^2=\eta^2=0$.
The Wigner function which follows from this has been derived independently
by Abe \cite{Abe}.\\
We easily get
\begin{eqnarray}
    a_W &=& i\theta\\
    (a^\dagger)_W &=&-i\eta
\end{eqnarray}
thus the conjugation of functions becomes
\begin{equation}
    (f(\theta,\eta))^* = \bar{f}(\eta,\theta)
\end{equation}
where the bar denotes Grassmann conjugation and the twisted product becomes
\begin{eqnarray}
    (f*g)(\theta,\eta) &=& 2(f_4g_4+3f_3g_2-f_2g_3+2f_1g_4+2f_4g_1) +
    \nonumber\\
    &&2(2f_1g_2-2f_2g_1-3f_2g_4-3f_4g_2)\theta+\nonumber\\
    && 2(2f_3g_1-2f_1g_3-f_3g_4-f_4g_3)\eta+\nonumber\\
    &&2(2f_4g_4-6f_3g_2+2f_2g_3)\theta\eta
\end{eqnarray}
where we have written $f=f_1+f_2\theta+f_3\eta+f_4\theta\eta$ and similar for
$g$. Contrasting this formula for the twisted product with the usual product
\begin{eqnarray*}
    (fg)(\theta,\eta) &=& f_1g_1+(f_1g_2+f_2g_1)\theta + (f_1g_3+f_3g_1)\eta
    +\\
    && (f_1g_4+f_4g_1+f_2g_3-f_3g_2)\theta\eta
\end{eqnarray*}
we see that the WWM-formalism introduces even more non-commutativity. With 
fermionic degrees of freedom within reach, the extension to super-Lie algebras
\cite{susy,DeWitt} is straightforward.

\subsection{Clifford and Spin Algebras}
I do not know of any concrete examples where the quantum phase-space is a
Clifford algebra, except of course the already treated case of ${\bf g}=su_2$.
Nevertheless it might be interesting to have a look at the WWM-formalism for
such algebras. Now, a Clifford algebra $C(r,s)$ is by definition an algebra in
$n=r+s$ generators $\gamma_a$ satisfying
\begin{equation}
    \{\gamma_a, \gamma_b\} = 2g_{ab}
\end{equation}
where $g_{ab}$ is a metric with signature $(r,s)$. We will simply assume
\begin{equation}
    g_{ab}=\eta_{ab} \equiv {\rm diag}(\underbrace{1,1,...,1}_r,\underbrace{-1,
    ,-1,...,-1}_s)
\end{equation} 
Note, that the definition implies $(\gamma_a)^2 = \pm 1$, hence $\dim C(r,s) =
2^{r+s}$. The case of $su_2$ corresponds to $r=2,s=0$ with $\gamma_1=\sigma_1,
\gamma_2=\sigma_2, \sigma_3=\frac{1}{2}\gamma_1\gamma_2$. 
Our ``classical coordinates'' $\xi_i$ will be taken to be G-numbers
anticommuting with the $\gamma$-matrices, $\{\xi_i,\gamma_j\}=0$. This 
would give a new representation of 
a classical phase-space of this algebra, in other words, $su_2$ as a Lie 
algebra must be treated differently from $su_2$ as a Clifford algebra.\\
Let me just sketch the results for the usual Clifford algebra $C(1,3)$, the
{\em Dirac algebra}. The translation operator is defined in the most natural 
way as
\begin{deff}
Let $\Gamma^I$ denote the generators of the Clifford algebra $C(r,s)$, then
\begin{equation}
    \Pi(\xi) = e^{i\xi_I\Gamma^I}
\end{equation}
where $\xi_I$ are G-numbers anticommuting with the Clifford generators.
\end{deff}
For $r=3,s=1$ -- the Dirac algebra -- we have
\begin{equation}
    \Pi(\xi) \equiv \exp(i\xi_01+i\tilde{\xi}_0\gamma_5
    +i\xi_m\gamma^m+i\tilde{\xi}_m\gamma_5\gamma^m+i\xi_{mn}\sigma^{mn})
\end{equation}
It has the decomposition (as do any function on a Clifford 
algebra)
\begin{equation}
    \Pi(\xi) = \Pi_0(\xi)+\tilde{\Pi}_0(\xi)\gamma^5 + \Pi_i(\xi)\gamma^i
    +\tilde{\Pi}_i(\xi)\gamma^i\gamma^5+\Pi_{ij}(\xi)\sigma^{ij}
\end{equation}
with
\begin{displaymath}
    \begin{array}{lclcl}
    \Pi_0(\xi) &\equiv & \frac{1}{4}{\rm Tr}\Pi(\xi) &\qquad & \mbox{(scalar)}\\
    \tilde{\Pi}_0(\xi) &\equiv & \frac{1}{4}{\rm Tr}(\Pi(\xi)\gamma^5) &&
    \mbox{(pseudoscalar)}\\
    \Pi_i(\xi) &\equiv & \frac{1}{4}{\rm Tr}(\Pi(\xi)\gamma_i) && \mbox{(vector)
    }\\
    \tilde{\Pi}_i(\xi) &\equiv & \frac{1}{4}{\rm Tr}(\Pi(\xi)\gamma_i\gamma_5)
    && \mbox{(axial vector)}\\
    \Pi_{ij}(\xi) &\equiv & \frac{1}{4}{\rm Tr}(\Pi(\xi)\sigma_{ij}) &&
    \mbox{(tensor)}
    \end{array}
\end{displaymath}
But, as the coefficients are G-numbers we have quite simply
\begin{eqnarray*}
    \Pi_0(\xi) &=& 1+i\xi_0\\
    \tilde{\Pi}_0(\xi) &=& \tilde{\xi}_0\\
    \Pi_m(\xi) &=& \xi_m\\
    \tilde{\Pi}_m(\xi) &=& \tilde{\xi}_m\\
    \Pi_{mn}(\xi) &=& \xi_{mn}
\end{eqnarray*}
Thus
\begin{equation}
    \left(\gamma_m\right)_W = \Pi_m(\xi) = i\xi_m
\end{equation}
while
\begin{equation}
    (1)_W = \Pi_0(\xi) = 1+i\xi_0
\end{equation}
and so on. It follows from this that it is natural to demand $\xi_0=0$, which
will lead to a dimensionality of $\Gamma$ of $\dim C(r,s)-1 =2^{r+s}-1$. Thus
\begin{prop}
Let $\Gamma$ denote the classical phase-space of a Clifford algebra $C(r,s)$
then
\begin{displaymath}
    \dim\Gamma = \dim C(r,s) -1 = 2^{r+s}-1
\end{displaymath}
as a Grassmann space.
\end{prop}
One should note that this always gives
an odd-dimensional space for any values of $r,s$. For the
Clifford algebra $su_2$ we thus have an alternative classical phase space,
namely a 3-dimensional Grassmann space. \\
The product of two ``translation'' operators is then
\begin{equation}
    \Sigma(\xi,\xi') \equiv \Pi(\xi)\Pi(\xi') = \Sigma_0+ \tilde{\Sigma}_0
    \gamma^5 + \Sigma_i\gamma^i+\tilde{\Sigma}_i\gamma^i\gamma^5 +
    \Sigma_{ij}\sigma^{ij}
\end{equation}
where
\begin{eqnarray}
    \Sigma_0 &=& \tilde{\xi}_0\tilde{\xi}'_0 
    -4i(\eta^{mp}\eta^{nq} - \eta^{mq}\eta^{np})\xi_{mn}\xi_{pq}'\\
    \tilde{\Sigma}_0&=& -4i\varepsilon^{mnpq}\xi_{mn} \xi_{pq}'\\
    \Sigma_m &=&-\tilde{\xi}_0 
    \tilde{\xi}_m'+\tilde{\xi}_m \tilde{\xi}_0'+\nonumber\\
    &&4i(\eta^{np}\delta^q_m-\eta^{nq}\delta^p_m)(\xi_n \xi_{pq}' +
    \xi_{pq} \xi_n')-4i\varepsilon^{npq}_{~~~m}(\tilde{\xi}_n 
    \xi_{pq}'+\xi_{pq}\tilde{\xi}_n')\\
    \tilde{\Sigma}_m &=&-\tilde{\xi}_0 \xi_m'+\xi_m \tilde{\xi}_0'
    -\nonumber\\
    && 4i\varepsilon^{npq}_{~~~m}(\xi_n \xi_{pq}'-\xi_{pq} 
    \xi_n')+4i(\eta^{np}\delta^q_m-\eta^{nq}\delta^p_m)(\tilde{\xi}_n 
    \xi_{pq}'+\xi_{pq} \tilde{\xi}_n')\\
    \Sigma_{mn} &=& 4i\varepsilon_{mn}^{~~pq}(\xi_{pq} \tilde{\xi}_0'
    +\tilde{\xi}_0 \xi_{pq}')+\nonumber\\
    && 4i(\eta^{pq}\eta^{rs}\eta_{mn}-\eta^{qr}\delta^s_m\delta^p_n +
    \eta^{rs}\delta^p_m\delta^q_n- \eta^{sp}\delta^q_m\delta^r_n) \xi_{pq} 
    \xi_{rs}'
\end{eqnarray}
The reproducing kernel $K(\xi,\xi')$ becomes
\begin{eqnarray}
    K(\xi,\xi') &\equiv & \frac{1}{4}{\rm Tr}(\Pi(\xi)\Pi(\xi')) =
    \frac{1}{4}\Sigma_0(\xi,\xi')\nonumber\\
        &=& \tilde{\xi}_0\tilde{\xi}'_0 
            -4i(\eta^{mp}\eta^{nq} -\eta^{mq}\eta^{np})\xi_{mn}\xi_{pq}'
\end{eqnarray}
While the kernel for the twisted products takes the form 
\begin{eqnarray}
    \Delta(\xi,\xi',\xi'') &\equiv & \frac{1}{4}{\rm Tr}(\Pi(\xi)\Pi(\xi')
    \Pi(\xi''))\nonumber\\
    &=& K(\xi,\xi')+K(\xi',\xi'')+K(-\xi,\xi'')
\end{eqnarray}
Now, any Clifford algebra can be written
\begin{eqnarray}
    C(r,s) &=& C_0(r,s)\oplus C_1(r,s)\oplus C_2(r,s)\oplus ... \oplus 
      C_n(r,s)\\
    &\equiv& C_{\rm even}(r,s)\oplus C_{\rm odd}(r,s)
\end{eqnarray}
where $C_k(r,s)$ consists of all powers of $k$ different generators, i.e. $C_0$
consists of the scalars, $C_1$ of the generators, $C_2$ of products of the form
$\gamma_i\gamma_j$ and so on, while $C_{\rm even}, C_{\rm odd}$ consists of all
linear combinations of products with an even an odd number of generators
respectively. To each such Clifford algebra two Lie groups are
defined, namely\footnote{The symbol $<C_1>$ denotes the group generated by all
the unit vectors in $C_1$, i.e. the group of products of generators $\gamma_i$.}
\begin{eqnarray}
    {\rm Pin(r,s)} &=& <C_1>\\
    {\rm Spin(r,s)} &=& {\rm Pin(r,s)}\cap C_{\rm even}(r,s)
\end{eqnarray}
and $Pin(r,s)$ is homomorphic to $O(r,s)$. It furthermore turns out that
the corresponding Lie algebra $spin(r,s)$ is ismorphic to $so(r,s)$, so we do
not get any new classical phase-spaces from that, even though the corresponding 
Lie groups $Spin(r,s)$ are inequivalent to any classical matrix group in all 
but a few cases, see \cite{GS}, as $Spin(r,s)$ is a covering group of $SO(r,s)$.
If, on the other hand, we do {\em not} consider $spin(r,s)$ as a classical Lie
algebra, but instead considers it as the Lie algebra of the non-classical Lie
group $Spin(r,s)$, which is built from the Clifford algebra $C(r,s)$, then we
{\em can} get new phase-spaces, namely Grassmann spaces. This leads, then, to an
alternative for the classical Lie algebras $so(r,s)$, as we have already seen
for $su_2=so(3)$. By construction, we must also have morphisms between the two
alternatives, the classical differentiable manifold $SO(r,s)/H$ and the
Grassmann spaces, thus allowing for the translation of problems of analysis on
$SO(r,s)/H$ into problems involving G-numbers, a possibility which should be of
quite some practical importance. 
One important difference is that, considering $so_{r,s}$ as a Lie algebra, we
get a symplectic manifold, whereas considering it as a Clifford algebra we get
an odd-dimensional Grassmann space.

\section{Quantum-Lie Algebras, Intermediate Statistics etc.}
We will make some very brief comments on the extension of the above method to
quantum groups. Given a (semisimple) Lie algebra $\bf g$ we can form its \
corresponding
{\em quantum universal algebra} $U_q({\bf g})$ \cite{Fuchs}, which is a 
deformed Lie 
algebra. A basis for this can be chosen in analogy with the ordinary Lie 
algebra case such that is satisfies
\begin{eqnarray*}
    \left[H^i, H^j\right] &=& 0\\
    \left[H^i, E^j_\pm \right] &=& \pm A^{ji}E^j_\pm\\
    \left[E^i_+, E^j_-\right] &=& \delta^{ij} \lfloor H^i\rfloor
\end{eqnarray*}
where the only new thing is the appearance of 
\begin{displaymath}
    \lfloor H^i\rfloor = \lfloor H^i\rfloor_q \equiv \frac{q^{\frac{1}{2}H^i} -
    q^{-\frac{1}{2}H^i}}{q^{\frac{1}{2}}-q^{-\frac{1}{2}}}
\end{displaymath}
on the right hand side above. It is here the quantum deformation $q$ enters. We
see that we can carry the formalism developed above for an arbitrary Lie 
algebra $\bf g$ over to its quantum universal algebra $U_q({\bf g})$ by making
the substitution
\begin{displaymath}
    H^i \rightarrow \lfloor H^i\rfloor
\end{displaymath}
in the definition of $Q(u,v;u',v')$ but not in $\Pi$. 
The logarithm og $Q$ would then be a highly
non-linear function of $H^i$ (it will be linear in $\lfloor H^i\rfloor$, 
though) and this non-linearity will be a measure of the deformation. The
corresponding quantum fibre bundle will now involve a double deformation of a
classical vector bundle. Would ``second quantized fibre bundle'' be a good 
name for such a structure? \\
We will just make some very brief comments on some further generalizations.
Bosons are described in terms of commutators and fermions in terms of
anti-commutators. Introducing the spin $s$ of the underlying field (integral 
for bosons, half-integral for fermions), we can write this as
\begin{equation}
    [a_k, a_l^\dagger ]_s \equiv a_k a_l^\dagger - (-1)^{2s+1} a_l^\dagger a_k =
    \delta_{kl}
\end{equation}
An obvious generalization is to allow $s$ to be any rational or even real
number, we can then define statistics interpolating between Bose-Einstein and
Fermi-Dirac statistics. Now, given two fermionic operators $a,a^\dagger$ we can
define bosonic ones by defining
\begin{displaymath}
    A = \alpha a \qquad\qquad  A^\dagger = \beta a^\dagger
\end{displaymath}
requiring that $(\alpha,\beta)$ are G-numbers which anticommute with the 
Fermi operators, we have $[A,A^\dagger] = \alpha\beta$, so when
$\beta=\bar{\alpha}$ and $\alpha$ is normalized to unity, then $A,A^\dagger$ 
are
ordinary Bose-operators. We can do a similar trick here by formally defining
``numbers'' which satisfy
\begin{displaymath}
    [\alpha,\beta]_s = 0\qquad\Rightarrow\qquad \alpha\beta = (-1)^{2s+1} \beta\alpha
\end{displaymath}
This will give us an ordinary Lie algebra in the formal operators $A_k,
A_k^\dagger$ and we know the WWM formalism for these, hence we can extend it to
these intermediate statistics aswell by using this little trick. The symplectic
product would then read
\begin{equation}
    (\alpha,\beta)\wedge(\alpha',\beta') = \alpha\beta'-(-1)^{2s+1}\beta\alpha'
\end{equation}
This leads to an alternative for quantum Lie algebras. If we have relations like
\begin{equation}    
    a_ka_l^\dagger = qR_{kl}^{~~k'l'}a_{l'}^\dagger a_{k'}
\end{equation}
then we need coordinates satisfying
\begin{eqnarray}
    x_ky_l &=& qR_{kl}^{~~k'l'}y_{l'}x_{k'}\\
    x_kx_l &=& x_lx_k\\
    y_ky_l &=& y_ly_k
\end{eqnarray}
So $\Gamma$ would become a {\em braided space} or a quantum-space. We can thus
establish morphisms between ordinary manifolds ($\bf g$ considered as a Lie
algebra, or $U_q({\bf g})$ considered as a deformation of $\bf g$), Grassmann
manifolds (${\bf g}=so(r,s)$ considered as a spin algebra) and braided spaces
($U_q({\bf g})$ considered as an algebra of transformations on such spaces). 
Such
morphism are of interest in their own right as they show relationships between
what would otherwise appear as unrelated areas of mathematics. \\
One could further consider general non-linear algebras, i.e. algebraic
structures satisfying
\begin{equation}
    \left[\lambda_i,\lambda_j\right] = iF_{ij}(\lambda)
\end{equation}
of which a quantum Lie algebra is but a particular case. As always, we will have
different options for the classical phase-space dependent upon how we interpret
this algebraic structure (i.e. as a deformation of an ordinary (super-)Lie
algebra, or as an algebra of automorphisms of some non-commutative structure
{\em \'{a} la} braided spaces). One could study parafermions and parabosons in
this way, for instance.

\section{Comment on Finite Groups}
All our emphasis so far has been on ``continous'' structures, Lie algebras
and structures derived therefrom, before we move on to discuss operator algebras
it is therefore appropriate to make a few comments on finite groups. Given a
finite group $G$, we can construct its algebra $C(G)$, this is the set of formal
linear combinations $\sum_{i=1}^{|G|}\alpha_i g_i$ with $\alpha_i\in{\bf {\sf F}}$ and
$G=\{g_i~|~i=1,...,n=|G|\}$. The coefficients $\alpha_i=\alpha(g_i)$ are thus
functions $G\rightarrow {\bf {\sf F}}$, and we can assume $G$ is a toplogical groups with
$\alpha_i$ continous, which explains the reason for the terminology 
$C(G)$.\footnote{The natural topology is the discrete one, of course, making 
all sets open and all functions continous.}\\
The idea is again, of course, to use
\begin{deff}
Let $G=\{e,g_1,...,g_{n-1}\}$ be a finite group, we define
\begin{equation}
    \Pi(u) = \exp(i\sum_{j=1}^{n'}u_jg_j-i\sum_{j=n'+1}^{n-1}\lambda_j(u)g_j)
\end{equation}
where $n=|G|, n'=|G|-|Z\backslash\{e\}|=|G|-|Z|+1$, with $Z$ denoting the 
center, and where we have
supposed $g_0=e$, the neutral element, which is not to be included as a proper
generator.
\end{deff}
This function $\Pi$ is
considered as a formal power series, and the coefficients $u_j,\lambda(u_j)$ can
in general be non-commutative (they are just formal quantities). In the case
where we have an identification of $G$ with a group of transformations over some
finite field (or division ring or even just prinicpal ideal domain), 
such as the {\em Chevalley groups}
$A_k({\bf {\sf F}}), B_k({\bf {\sf F}}), C_k({\bf {\sf F}}), D_k({\bf {\sf F}})$ which generalize the usual Lie algebras of
the same names, see \cite{Carter}, it would be natural to let $u_j,\lambda_j$ 
belong to this finite field (or division ring) ${\bf {\sf F}}$. \\
Thus there is an ambiguity in the definition for finite groups, as we have no
{\em a priori} candidate for ${\bf {\sf F}}$, the field (or even just ring) to which the
coefficients in the algebra $C(G)$ of $G$ belongs. Choosing an infinite field
like ${\bf {\sf F}}={\bf {\sf R}}$ or ${\bf {\sf F}}={\bf {\sf C}}$ would just give us ordinary Lie algebras, whereas
infinite field such as ${\bf {\sf Q}}, {\bf {\sf Q}}(\alpha_1,...,\alpha_n)$, with $\alpha_i$
transcendent over ${\bf {\sf Q}}$, would lead to something slightly different, of use,
perhaps, in Galois theory, while choosing a finite field ${\bf {\sf F}}={\bf {\sf Z}}_p={\bf {\sf Z}}/p{\bf {\sf Z}}$, $p$ 
a prime,
or ${\bf {\sf F}}=GF(p^n)$, (the so-called {\em Galois field}), 
would lead to something very different, namely a finite, discrete set (i.e. a
kind of lattice) as the classical phase-space.\\
Let us furthermore notice that for finite groups we have $g^n=e$ for any element
$g$ of the group, and so the exponential is well-defined, and can in fact be
``decomposed'' as
\begin{equation}
    \Pi(u) = 1+\sum_{j=0}^{n-1} \pi_j(u)g_j
\end{equation}
For the cases $su_2,su_3$ and Clifford algebras we had a similar decomposition
wich was very useful for practical calculations. The functions $\pi_j(u)$ are
Taylor series if the field has characteristic zero, and polynomials otherwise.\\
Before we look at some examples let us notice that the phase-space of a Galois
extension ${\bf {\sf F}}(\alpha)$ can be obtained from that of the original field ${\bf {\sf F}}$ in a
simple manner. Let ${\bf {\sf F}}(\alpha)$ have dimension $n$ as a vector space over ${\bf {\sf F}}$,
i.e. $|{\bf {\sf F}}(\alpha):{\bf {\sf F}}|=n$, then ${\bf {\sf F}}(\alpha) = {\bf {\sf F}}\oplus \alpha{\bf {\sf F}}\oplus...\oplus
\alpha^{n-1}{\bf {\sf F}}$, so any element in the Galois extension can be written as
$u=u_0+u_1\alpha+...+u_{n-1}\alpha^{n-1}$. So the transition ${\bf {\sf F}}\rightarrow
{\bf {\sf F}}(\alpha)$ can be written $u\mapsto u(\alpha)=u_0+u_1\alpha+...+u_{n-1}
\alpha^{n-1}$. We have thus proven
\begin{prop}
Let ${\bf {\sf F}}$ be any field and let $\alpha$ be transcendent over ${\bf {\sf F}}$, for any
Chevalley algebra $\bf g$ over ${\bf {\sf F}}$ we then have
\begin{equation}
    \Gamma_{\bf g}({\bf {\sf F}}(\alpha)) = \Gamma_{\bf g}({\bf {\sf F}})\otimes{\bf {\sf F}}(\alpha)
\end{equation}
\end{prop}
A result very similar to the ones for loop algebras or complexifications we 
saw earlier.

\subsection{Examples of Finite Groups}
To develop the formalism I will just give a two examples the permutation 
group $S_3$ and the Chevalley group $A_1({\bf {\sf F}})$, ${\bf {\sf F}}$ some field (finite or
infinite).\\
For the permutation groups $S_3, A_3$ we have the multiplication table as shown
in table 2, with $A_3$ being the subgroup made up by $\{e,g_1,g_4,\}$, 
which is also the largest abelian subgroup. From this we get
\begin{eqnarray}
    \Pi(u) &\equiv & e^{-i\lambda_1 g_1+iu_2g_2+iu_3g_3-i\lambda_2g_4+iu_5g_5}\\
    &=& 1+\pi_0(u)e+\sum_{i=1}^5\pi_i(u)g_i
\end{eqnarray}
where
\begin{equation}
    \pi_i(u) = \sum_{n=1}^\infty \frac{i^n}{n!}\alpha_i^{(n)} \qquad i=0,1,...,5
\end{equation}
with the coefficients $\alpha_i^{(n)}$ given by the recursion relations
\begin{eqnarray}
    \hspace{-7mm}\alpha_0^{(n+1)} &=& \alpha_0^{(n)}\alpha_0^{(1)}+\alpha_1^{(n)
    }\alpha_4^{(1)}+\alpha_2^{(n)}\alpha_2^{(1)}+\alpha_3^{(n)}\alpha_3^{(1)}
    +\alpha_4^{(n)}\alpha_1^{(1)}+\alpha_5^{(n)}\alpha_5^{(1)}\\
    \alpha_1^{(n+1)} &=& \alpha_0^{(n)}\alpha_1^{(1)}+\alpha_1^{(n)
    }\alpha_0^{(1)}+\alpha_2^{(n)}\alpha_5^{(1)}+\alpha_3^{(n)}\alpha_2^{(1)}
    +\alpha_4^{(n)}\alpha_4^{(1)}+\alpha_5^{(n)}\alpha_3^{(1)}\\
    \alpha_2^{(n+1)} &=& \alpha_0^{(n)}\alpha_2^{(1)}+\alpha_1^{(n)
    }\alpha_5^{(1)}+\alpha_2^{(n)}\alpha_0^{(1)}+\alpha_3^{(n)}\alpha_1^{(1)}
    +\alpha_4^{(n)}\alpha_3^{(1)}+\alpha_5^{(n)}\alpha_4^{(1)}\\
    \alpha_3^{(n+1)} &=& \alpha_0^{(n)}\alpha_3^{(1)}+\alpha_1^{(n)
    }\alpha_2^{(1)}+\alpha_2^{(n)}\alpha_4^{(1)}+\alpha_3^{(n)}\alpha_0^{(1)}
    +\alpha_4^{(n)}\alpha_5^{(1)}+\alpha_5^{(n)}\alpha_1^{(1)}\\
    \alpha_4^{(n+1)} &=& \alpha_0^{(n)}\alpha_4^{(1)}+\alpha_1^{(n)
    }\alpha_1^{(1)}+\alpha_2^{(n)}\alpha_3^{(1)}+\alpha_3^{(n)}\alpha_5^{(1)}
    +\alpha_4^{(n)}\alpha_0^{(1)}+\alpha_5^{(n)}\alpha_2^{(1)}\\
    \alpha_5^{(n+1)} &=& \alpha_0^{(n)}\alpha_5^{(1)}+\alpha_1^{(n)
    }\alpha_3^{(1)}+\alpha_2^{(n)}\alpha_1^{(1)}+\alpha_3^{(n)}\alpha_4^{(1)}
    +\alpha_4^{(n)}\alpha_2^{(1)}+\alpha_5^{(n)}\alpha_0^{(1)}
\end{eqnarray}
subject to
\begin{equation}
    \alpha_0^{(1)} = 0\qquad \alpha_i^{(1)}=u_i\mbox{ for }i=2,3,5\mbox{  and  }
    \alpha_1^{(1)} = -\lambda_1\qquad\alpha_4^{(1)}=-\lambda_2
\end{equation}
The dimensionality of the ``phase-space'' (with a field of characteristic zero
as underlying field) is then $|G|-|A_3| = 6-3=3$. The deformed addition is
rather complicated, namely
\begin{equation}
    \left(\begin{array}{c}u_2\\ u_3\\ u_5\end{array}\right)\oplus
    \left(\begin{array}{c}u_2'\\ u_3'\\ u_5'\end{array}\right) =
	\left(\begin{array}{c} u_2+u_2'-u_3\lambda_1'-\lambda_2 u_3'-u_5
	\lambda_2'-\lambda_1u_5'\\
	u_3+u_3'-\lambda_2u_2'-u_2\lambda_2'-\lambda_2u_5'-u_5\lambda_1'\\
	u_5+u_5'-\lambda_1u_3'-u_2\lambda_1'-u_3\lambda_2'-\lambda_2u_2'
	\end{array}\right)
\end{equation}
The ``undeformed'', or ``zero'th order'' antisymmetric two-form $\omega_0$ is
the
coefficient, to the lowest order, of the Cartan elements, hence (for a general
Lie algebra, with root-decomposition as in the text)
\begin{equation}
    \omega_0(u,v,u',v') = \sum_\alpha (u_\alpha v_\alpha'-u_\alpha' v_\alpha)
\end{equation}
this is then the analogue of the Poisson bracket when $\dim\Gamma$ is even.
For our case it is similarly
\begin{eqnarray}
    \omega_0(u,u')&=& u_2u_5'-u_2'u_5+u_3u_2'-u_3'u_2+u_5u_3'-u_5'u_3\nonumber\\
    &=&\left|\begin{array}{ccc}
        -1 & u_2 & u_2'\\ -1 & u_3 & u_3'\\ -1 & u_5 & u_5'
    \end{array}\right|
\end{eqnarray}
The Chevalley group of $A_1({\bf {\sf F}})$ over any field (finite or infinite), ${\bf {\sf F}}$, is 
defined from the relations
\begin{equation}
    \left[e,f\right]=h\qquad \left[h,e\right]=e\qquad\left[h,f\right]=-f
\end{equation}
letting $A_1({\bf {\sf Z}})$ denote the ${\bf {\sf Z}}$-linear span of these elements we get a Lie
algebra, for any field ${\bf {\sf F}}$ we then put
\begin{equation}
    A_1({\bf {\sf F}}) \equiv A_1({\bf {\sf Z}})\otimes{\bf {\sf F}}
\end{equation}
For ${\bf {\sf F}}={\bf {\sf R}}$ we get $sl_2({\bf {\sf R}})=so_3=su_2$ whereas for ${\bf {\sf F}}={\bf {\sf C}}$ we get their
respective
complexifications. For a finite field ${\bf {\sf F}}=GF(p^n)$ (with $GF(p)={\bf {\sf Z}}_p$) we get
something completely new, and for ${\bf {\sf F}}={\bf {\sf Q}}$ we get $sl_2({\bf {\sf Q}})$. Let us 
concentrate upon
${\bf {\sf F}}={\bf {\sf Z}}_p$ for now. The phase-space cannot simply, as for the infinite fields
${\bf {\sf R}},{\bf {\sf C}}$, be diffeomorphic to $\{x,y,z\in{\bf {\sf F}}~|~x^2+y^2+z^2=1\}$ as spheres of
different radii will contain an unequal number of points in the discrete case.
\\
The subgroup $H$ is just the diagonal subgroup, and hence is isomorphic to
${\bf {\sf F}}^\times$, where ${\bf {\sf F}}^\times$ denotes the set of invertible elements in ${\bf {\sf F}}$
(for ${\bf {\sf F}}$ a field and not just a division ring, this is ${\bf {\sf F}}\backslash\{0\}$).
Hence, since the group with Lie algebra $A_1({\bf {\sf F}})$ is $PSL_2({\bf {\sf F}})$ (see Carter
\cite{Carter})
\begin{equation}    
    \Gamma_{A_1}({\bf {\sf F}}) \simeq PSL_2({\bf {\sf F}})/{\bf {\sf F}}^\times
\end{equation}
For an infinite field such as ${\bf {\sf Q}}$ or one of its Galois extensions, this is a
``manifold'' of dimension 2, as for ${\bf {\sf F}}={\bf {\sf R}},{\bf {\sf C}}$, whereas for finite fields it is
a finite set of points. For ${\bf {\sf F}}=GF(p^n)$ for some prime $p$ and some integer
$n$, we have
\begin{equation}
    |\Gamma| = \frac{1}{(2,p^n-1)}p^{2n}(p^{2n}-1)-(p^n-1)
\end{equation}
where we have used $|GF(p^n)|=p^n$ and where $(a,b)$ denotes the greatest common
divisor of $a,b$. In the special case $n=1$, in which case $GF(p)\simeq{\bf {\sf Z}}_p$, we
thus get a set consisting of $11$ points for $p=2$, $34$ for $p=3$ and so on.\\
I will leave the discussion of finite groups at this point to give a summary of
properties derived so far, and then go on to operator algebras. The further
development of a WWM-formalism for finite groups will certainly be of interest
in its own right (applications to pure algebra, Galois theory and algebraic
geometry spring to mind), but I do not know of any physical situation which 
could serve as a motivation.

\section{Summary of Properties}
We will finish off this discussion with a summary of the algebraic properties 
of
the WWM-formalism we have been developing. The formalism consists basically of
(1) $\Pi$ and $Q$, the maps defining the Weyl transformation and its algebraic
properties, (2) the set $C(\Gamma)$ of functions $\Gamma\rightarrow {\bf {\sf C}}$, where
$\Gamma$ is the classical phase-space. The basic correspondence is
\begin{eqnarray*}
    A_W(\xi) &\equiv& {\rm Tr}~\Pi(\xi)\hat{A}\\
    \hat{A} &\equiv & \int_\Gamma \Pi(\xi)A_W(\xi) d\mu
\end{eqnarray*}
where the Weyl transform $\hat{A}\mapsto A_W$ is an isomorphism $U({\bf g})
\rightarrow C(\Gamma)$.\footnote{This only holds, of course, for ``good''
algebras, such as for instance semisimple or abelian, in general we might only
have a homomorphism.} The operator-valued function $\Pi$ can be viewed as a
``translation'' operator and satisfies
\begin{displaymath}
    \Pi(\xi)\Pi(\xi') = \Pi(\xi\oplus \xi') Q(\xi\times \xi')
\end{displaymath}
The operations $\oplus, \times$ were referred to as the deformed addition and
symplectic product respectively. For an abelian algebra $\xi\oplus \xi' =
\xi+\xi'$ and thus the deformation is a measure of the non-commutativity.
Furthermore, the classical phase-space $\Gamma$ is a vector space if the 
algebra
is abelian and a symplectic manifold if $\bf g$ is semisimple or obtained from
a semisimple Lie algebra by a central extension or by adding an 
abelian algebra. Its dimensionality is
\begin{displaymath}
    \dim \Gamma = \dim~{\bf g}-{\rm rank}~{\bf g}\equiv n-l
\end{displaymath}
and for $n=\dim{\bf g}<\infty$ we have
\begin{displaymath}
    \Gamma_{\bf g} = G/H
\end{displaymath}
where $G$ is the smallest connected Lie group having $\bf g$ as its Lie 
algebra, while $H$ is
similar but for the Cartan subalgebra of $\bf g$.\\
We discovered some very nice properties of $(\Pi, Q, \Gamma)$, namely
\begin{eqnarray*}
    {\bf g} = {\bf g}_1\oplus{\bf g}_2 &\Rightarrow& \Pi_{\bf g} = \Pi_{{\bf g}_1}
    \Pi_{{\bf g}_2}\mbox{  and  } Q_{\bf g} = Q_{{\bf g}_1}Q_{{\bf g}_2}\\
    {\bf g} = {\bf g}_1+{\bf g}_2 &\Rightarrow & \Pi_{\bf g} = \Pi_{{\bf g}_1}
    \Pi_{{\bf g}_2}\mbox{  and  } Q_{\bf g} = Q_{{\bf g}_1}Q_{{\bf g}_2}q_Z
    \mbox{  if  } \left[{\bf g}_1, {\bf g}_2\right] \subseteq Z({\bf g})\\
    {\bf h}\mbox{ ideal in }{\bf g} &\Rightarrow & \Pi_{\bf g} =\Pi_{{\bf g}/{\bf h}}
    \Pi_{\bf h}\mbox{  and  } Q_{\bf g} = Q_{{\bf g}/{\bf h}}Q_{\bf h}
\end{eqnarray*}
which allows us to study central extension very easily (for instance to express
the WWM-formalism for an affine Kac-Moody algebra in terms of the WWM-formalism
for a loop algebra). Another very important property was
\begin{displaymath}
    \Gamma({\bf g}\otimes C^\infty(M)) =\Gamma(C^\infty(M\rightarrow {\bf g})) 
    \simeq \Gamma({\bf g})\otimes C^\infty(M) = C^\infty(M\rightarrow\Gamma
    ({\bf g}))
\end{displaymath}
which allows to gauge an algebra and extend or WWM-formalism easily, in
particular we can go to the loop algebra $M=S^1$. A similar results hold for
Galois extensions of the base field ${\bf {\sf F}}\rightarrow {\bf {\sf F}}(\alpha_1,...,\alpha_n)$
\begin{displaymath}
    \Gamma_{\bf g}({\bf {\sf F}}(\alpha_1,...,\alpha_n)) \simeq \Gamma_{\bf g} \otimes
    {\bf {\sf F}}(\alpha_1,...,\alpha_n)
\end{displaymath}
For ${\bf {\sf F}}={\bf {\sf R}}, \alpha=\pm i$ we get a result about complexifications.\\
A final result relates to morphisms $\phi:{\bf g}_1\rightarrow {\bf g}_2$, i.e.
structure-preserving maps between algebras (homomorphisms for Lie
algebras; Jordan maps, i.e linear maps preserving the anticommutator, for 
fermions; super-Lie homomorphisms for super-Lie algebras and so on). Any such
morphism induces a map $\Phi: C(\Gamma_1) \rightarrow C(\Gamma_2)$ where
$\Gamma_i$ is the phase-space of ${\bf g}_i$. Consider the commutative diagram
\begin{displaymath}
    \begin{array}{ccc} U({\bf g}_1) & \stackrel{\phi}{\longrightarrow}&
                U({\bf g}_2)\\
             \left.\begin{array}{c} ~\\{\scriptsize\Pi_1}\\~
                    \end{array}\right\downarrow & &
             \left\downarrow\begin{array}{c} ~\\{\scriptsize\Pi_2}\\~
                    \end{array}\right.\\
                C(\Gamma_1) & \stackrel{\Phi}{\longrightarrow}& C(\Gamma_2)
    \end{array}
\end{displaymath}
using that $\Pi_1$ is an isomorphism we can define
\begin{displaymath}
    \Phi = \Pi_2 \circ\phi \circ\Pi_1^{-1}
\end{displaymath}
and then $\Phi$ is well defined and unique.\\
We can use this to carry topological and algebraic structure form $\bf g$
through $U({\bf g})$ to
$C(\Gamma)$. Suppose for instance that $\bf g$ is a normed or semi-normed 
space,
i.e. it is endowed with a map $\rho: {\bf g}\rightarrow {\bf {\sf R}}$ which is sublinear
($\rho(A+B) \leq \rho(A)+\rho(B)$) and positive homogenous ($\rho(\alpha A) =
|\alpha|\rho(A)$ with $\alpha$ a scalar). Noting that $\Gamma({\bf {\sf R}}) =\{0\}$, i.e.
$C(\Gamma({\bf {\sf R}})) \simeq {\bf {\sf R}}$ (similar for ${\bf {\sf C}}$, of course)
we have the commutative diagram 
\begin{displaymath}
    \begin{array}{ccc}
        {\bf g} & \stackrel{\rho}{\rightarrow} & {\bf {\sf R}}\\
        \downarrow & & \downarrow\\
        C(\Gamma) &\stackrel{\tilde{\rho}}{\rightarrow} & C(\Gamma({\bf {\sf R}}))\simeq{\bf {\sf R}}
    \end{array}
\end{displaymath}
thus $C(\Gamma)$ is a normed or semi-normed space whenever $\bf g$ is. Hence
$C(\Gamma)$ is a Banach space if and only if $\bf g$ is, and the mapping $\Pi$
becomes an isometry in this case. Similarly, if $\bf g$ comes equipped with an
inner product, i.e. a sesquilinear map ${\bf g}\times{\bf g}\rightarrow {\bf {\sf C}}$,
then $\Pi$ induces a sesquilinear form on $C(\Gamma)$, which then 
becomes Hilbert if and only if $\bf g$ is a Hilbert space. The diagram is
\begin{displaymath}
    \begin{array}{ccccc}
        {\bf g} &\rightarrow &{\bf g}\times{\bf g} &\rightarrow &{\bf {\sf C}}\\
        \downarrow & & \downarrow &&\downarrow\\
        C(\Gamma) & \rightarrow &C(\Gamma)\times C(\Gamma) & \rightarrow & {\bf {\sf C}}
    \end{array}
\end{displaymath}
We should note that semisimple Lie algebras come with a natural non-degenerate
bilinear form and will thus give pre-Hilbert spaces.\\
Let us also note that this shows that our construction is in fact independent 
of the representation: considering ${\bf g}_1, {\bf g}_2$ to be two faithfull
irreducible representations of a given Lie algebra ${\bf g}$, i.e. we have
isomorphisms $\rho_i:
{\bf g}\rightarrow {\bf g}_i\subseteq gl_{n_i}$, this induces an isomorphism 
${\bf g}_1\rightarrow
{\bf g}_2$ and hence their two classical phase-spaces will be equivalent. The
diagram is
\begin{displaymath}
    \begin{array}{ccc}
    U({\bf g}_1) & \stackrel{\rho_2\circ\rho_1^{-1}}{\longrightarrow} &
    U({\bf g}_2)\\
    \left.\begin{array}{c}~\\ ~\\{\scriptsize \Pi_1}\\ ~\\ \end{array}
    \right\downarrow & \begin{array}{ccc} \stackrel{\rho_1}{\nwarrow} & &
    \stackrel{\rho_2}{\nearrow}\\
        & U({\bf g}) &\\
        & \downarrow {\scriptsize \Pi} & \\
        & C(\Gamma) &\\
    \stackrel{\tilde{\rho}_1}{\swarrow} & & \stackrel{\tilde{\rho}_2}{\searrow}
        \end{array} & \left\downarrow\begin{array}{c}~\\ ~\\{\scriptsize
        \Pi_2}\\ ~\\ \end{array}\right.\\
    C(\Gamma_1) &\stackrel{\widetilde{\rho_2\circ\rho_1^{-1}}}{\longrightarrow}
    & C(\Gamma_2)
    \end{array}
\end{displaymath}
with
\begin{equation}
    \widetilde{\rho_2\circ\rho_1^{-1}} = \tilde{\rho}_2\circ\tilde{\rho}_1^{-1}
\end{equation}
Furthermore, any diffeomorphism $\alpha: \Gamma_1\rightarrow \Gamma_2$
induces a map $\alpha_*: C(\Gamma_1)\rightarrow C(\Gamma_2)$, which then 
leads to
a map $\tilde{\alpha}: U({\bf g}_1)\rightarrow U({\bf g}_2)$, which allows us
to study the group of maps $\alpha$ of one manifold onto another in a new,
more algebraic way.\\
We have
\begin{prop}
If $\bf g$ is a normed algebra then so is $C(\Gamma)$, if $\bf g$ has an inner
product then so does $C(\Gamma)$. Thus if $\bf g$ is Hilbert or Banach, then so
is $C(\Gamma)$.
\end{prop}
All of the above holds for a very large class of algebraic structures as we 
have seen.

\section{$C^*$-Algebras}
It would be interesting to go on to an even larger
class of algebras such as $C^*$-algebras. The general idea is to
construct an isomorphism
\begin{displaymath}
    {\cal A} \rightarrow C(\Gamma)
\end{displaymath}
between a $C^*$-algebra and an algebra of functions on some manifold 
$\Gamma$. For
{\em abelian} algebras such an isomorphism is already known (the Gel'fand 
theorem \cite{BR,C*})
\begin{displaymath}
    {\cal A} \simeq C_0(X)
\end{displaymath}
where $C_0$ denotes the functions vanishing at infinity and $X$ is some locally
compact Hausdorff space (the spectrum or maximal ideal space of $A$)
which is compact if and only if $\cal A$ contains the
identity, \cite{BR,C*}. Our WWM formalism would then provide us with a {\em 
non-abelian Gel'fand theorem}. One
should note that the basic ingredient in Gel'fand's theorem is the concept
of a {\em character} on an abelian $C^*$-algebra, i.e. a linear map $\chi: 
{\cal A} \rightarrow {\bf {\sf C}}$ such that $\chi(AB)=\chi(A)\chi(B)$, $X$ is the space
of such maps, and is hence a subset of the dual ${\cal A}^*$ of $\cal A$. 
The WWM formalism
gives a natural generalization of this: $\chi(A)=A_W$, the product rule then
reads $\chi(AB)=\chi(A)*\chi(B)$ and we could refer to the Weyl transform as
a {\em generalized character}. The major
problem is the construction of $\Gamma$ (the abelian case uses ${\cal A} 
\subset
{\cal A}^{**}$ and $X\subset {\cal A}^*$, hence $\cal A$ can be viewed as
functions on $X$, it then relies on the Stone-Weierstrass theorem to prove the
isomorphism, and this is difficult to generalize to non-abelian algebras).\\
Any non-abelian $C^*$-algebra is isomorphic to a subalgebra of the algebra
${\cal B}({\cal H})$ of bounded operators on some seperable Hilbert space $\cal
H$. The method developed in the previous
sections can thus be seen as a special case, namely the case of finite
dimensional $C^*$-algebras, and we now want to go further.
A particular important subalgebra $\cal B$ is
${\cal K} ={\cal B}_0({\cal H})$ of {\em compact operators}, i.e. the operators
for which the image of the unit ball $\{x\in {\cal H}~|~ \|x\|^2 \leq 1\}$ is
compact. The elements of this subalgebra can be approximated by finite
matrices, in fact \cite{Kth,C*}
\begin{displaymath}
    {\cal K} = \lim_\rightarrow gl_n({\bf {\sf C}})
\end{displaymath}
where the $\lim_\rightarrow$ is understood as the {\em inductive limit}, hence
$\cal K$ is the completion (in norm-topology) of $gl_\infty({\bf {\sf C}})$. This
suggest that the case of compact operators is the next simplest case to 
treat.\footnote{A $C^*$-algebra which can be obtained as the inductive limit of
matrix algebras is known as an {\em AF-algebra}, an ``approximately finite
dimensional'' algebra. Thus our methods can be generalized to these.} And
in fact we can use the very definition of inductive limit to construct directly
the corresponding classical phase-space. Recall that the inductive limit
requires a {\em directed system} $\{A_i, \Phi_{ij}\}_{i\in {\cal I}}$, i.e. a
family of objects $A_i$ indexed by an upward filtering index set $\cal I$
(i.e. a set $\cal I$ such that whenever
$i,j \in {\cal I}$ a $k\in {\cal I}$ exists such that $k > i$ and $k>j$) and
with a morphism $\Phi_{ij}:A_j\rightarrow A_i$ whenever $j>i$. The inductive
limit $A_\infty$ is then the object $\bigcup_{\cal I} A_i$ with morphisms
$\Phi_i: A_i\rightarrow A_\infty$ such that
\begin{displaymath}
    \begin{array}{ccc}
        A_j & \stackrel{\Phi_j}{\rightarrow }& A_\infty\\
        \left.\begin{array}{c} ~\\{\scriptsize\Phi_{ij}}\\~
              \end{array}\right\downarrow
        & \stackrel{\Phi_i}{\nearrow}&\\
        A_j&&
    \end{array}
\end{displaymath}
commutes.\\
Denoting by $\Pi_n$ the WWM-map from $M_n=gl_n({\bf {\sf C}})$ into $C(\Gamma_n)$, where
$\Gamma_n$ is the classical phase-space corresponding to $M_n$, we get the
following diagram
\begin{displaymath}
  \begin{array}{ccc}
    M_n & \stackrel{\Phi_{mn}}{\longrightarrow} & M_m\\
    \left.\begin{array}{c} ~\\ \\{\scriptsize\Pi_n}\\ \\ 
        \end{array}\right\downarrow &
    \begin{array}{ccc} \stackrel{\Phi_n}{\searrow} && \stackrel{\Phi_m}
    {\swarrow}\\
    & M_\infty & \\
    & \downarrow \Pi_\infty &\\
    & C(\Gamma_\infty)& \\
    \stackrel{\tilde{\Phi}_n}{\nearrow} && \stackrel{\tilde{\Phi}_m}{\nwarrow}
    \end{array} &
    \left\downarrow\begin{array}{c} ~\\ \\{\scriptsize\Pi_m}\\ \\ 
    \end{array}\right.\\
    C(\Gamma_n) & \stackrel{\tilde{\Phi}_{mn}}{\longrightarrow} & C(\Gamma_m)
  \end{array}
\end{displaymath}    
Expressed in formulas we have
\begin{equation}
    C(\Gamma({\cal K})) = C(\Gamma_\infty)) = \lim_\rightarrow C(\Gamma_n)
\end{equation}
The map $\Pi_\infty$ is given by
\begin{equation}
    \Pi_\infty (A) \equiv \lim_{n\rightarrow\infty} \Pi_n(P_nAP_n)
\end{equation}
where $\Pi_n$ is, as in the diagram, the Weyl map for $gl_n$ and where $P_n$ is
the projection ${\cal K}\rightarrow gl_n$, these constitute an approximate unit
for $\cal K$ (i.e. $P_nA\rightarrow A~~\forall A\in{\cal K}$)
and the above construction is then well-defined.\\
If we could extend our scheme to ${\cal B}({\cal H})$ then we were able to 
treat
any $C^*$-algebra, thus our next problem is to find out how to go from ${\cal
K}= {\cal B}_0$ to ${\cal B}$. One way is to write down an exact 
sequence\footnote{A
sequence $A\stackrel{\alpha}{\rightarrow}B\stackrel{\beta}{\rightarrow}C$ is
said to be {\em exact} if the kernel of $\beta$ is the image of $\alpha$, i.e.
going twice ($\beta\circ\alpha$) gives zero, and this is the only way of 
getting zero. Hence $0\rightarrow A\stackrel{\alpha}{\rightarrow}B$ is 
exact if and only
if $\alpha$ is injective, and $A\stackrel{\alpha}{\rightarrow}B\rightarrow 0$ 
is
exact if and only if $\alpha$ is surjective. This notion is easily generalized
to longer sequences, we simply demand the kernel of one map to be equal to the
image of the previous one.}
\begin{displaymath}
    0\rightarrow {\cal K}\rightarrow {\cal B}\rightarrow {\cal B}/{\cal K}
    \rightarrow 0
\end{displaymath}
where ${\cal B}/{\cal K}$ is known as the {\em Calkin algebra}, this shows that 
${\cal B}$ is an {\em extension} of the algebra ${\cal K}$ by the 
Calkin algebra. There is
another way of obtaining $\cal B$ from  $\cal K$, namely by the use of what is
known as the {\em multiplier algebra} ${\cal M}(A)$ of a $C^*$-algebra, this is
defined as the largest unitization of $A$\footnote{i.e. the largest algebra
constructed from $A$ containing $A$ itself and a unit element $1$}, 
and can be constructed as follows.
Suppose $A$ acts non-degeneratly on some Hilbert space ${\cal H}_1$ (this
is always possible to arrange), then $A\subseteq {\cal B}({\cal H}_1)$ and we 
put
\begin{equation}
    {\cal M}(A) = \{x\in {\cal B}({\cal H}_1)~|~ xA\subseteq A ~\wedge~ Ax
    \subseteq A\}
\end{equation}
equivalently, ${\cal M}(A)$ is the completion in the topology induced by the
seminorms $x\mapsto \|xa\|$ and $x\mapsto \|ax\|$ where $x\in {\cal B}({\cal
H}_1)$ and $a\in A$ (this topology is known as the {\em strict topology}). 
The basic result is
\begin{equation}
    {\cal M}({\cal K})  = {\cal B}
\end{equation}
Thus if we can find a way of extending the WWM-formalism for a given
$C^*$-algebra $A$ consisting of compact operators ($A\subseteq {\cal K}$), to
its multiplier algebra ${\cal M}(A)$ then we have extended our WWM-formalism to
all $C^*$-algebras. Another interesting possibility, closely related to this, 
is
the study of the WWM-formalism for arbitrary extensions of $A$. This would also
be an interesting excercise in the case of Lie algebras, as would the study of
In\"{o}n\"{u}-Wigner contractions.\\
Before doing this let us look at the simplest (smallest) unitization $A^+$ of
$A$, when $A$ is not itself unital then $A^+\simeq A+1{\bf {\sf C}}$, i.e. $x=a+\lambda,
x\in A^+, a\in A, \lambda \in{\bf {\sf C}}$ with a natural product $(a+\lambda)(b+\mu) =
ab+\lambda b+\mu a + \lambda\mu$. Any morphism $\phi:A\rightarrow B$ between
$C^*$-algebras induces a morphism $\phi^+:A^+\rightarrow B^+$ given by
\begin{displaymath}
    \phi^+(a+\lambda) \equiv \phi(a)+\lambda
\end{displaymath}
Letting $B=C(\Gamma)$ and $\phi=\Pi$ we get
\begin{equation}
    C(\Gamma^+) = C(\Gamma(A^+)) \simeq C(\Gamma)\times{\bf {\sf C}}
\end{equation}
any function in $C(\Gamma^+)$ is thus a pair $(f(x),\lambda)$ where $f:
A\rightarrow {\bf {\sf C}}$ and $\lambda\in{\bf {\sf C}}$.
This implies that $\Gamma(A^+)\equiv \Gamma^+$ is constructed by the 
adjoining of a
point to $\Gamma(A)=\Gamma$; the scalar $\lambda$ is then the value assigned
to $f$ at this extra point, i.e. we can consider $\Gamma^+$ to be the one-point
compactification of $\Gamma$, in standard symbols:
\begin{equation}
    \Gamma^+ = \alpha \Gamma
\end{equation}
For $C^*$-algebras the adjoining of a unit does not lead to the old 
phase-space plus some isolated point, as we always have sequences 
$e_n\rightarrow 1~,~ e_n\in A$ (approximate units), so the new phase-space, 
which is again the old one with some point added, must be just as
connected as the original one, thus leading to a compactification as argued 
above. For Lie algebras we do not have any sequences corresponding to 
approximate units, and hence get isolated points.\\
Now, the Gel'fand theory for abelian $C^*$-algebras give exactly this
relationship too, which seems to imply that our scheme is indeed in some sense
the non-commutative version of Gel'fand's. Similarly we can see that any 
unitization of $A$ leads to a compactification of $\Gamma$:
\begin{displaymath}
    \mbox{ unitization of }A \longrightarrow \mbox{ compactification of }\Gamma
\end{displaymath}
Let $A_1, A_2$ be two different unitizations of $A$, then $A_1\subseteq A_2$
implies $\Gamma_1\subseteq\Gamma_2$, where $\Gamma_i = \Gamma(A_i)$. Now, the
smallest unitization should thus correspond to the smallest compactification
(which we also saw that it did) and the largest unitization, the multiplier
algebra ${\cal M}(A)$, to the largest compactification $\beta\Gamma$, the
Stone-\v{C}ech compactification. Thus
\begin{equation}
    \Gamma_{{\cal M}(A)} = \beta\Gamma_A
\end{equation}
and the {\em corona algebra} ${\cal M}(A)/A$ becomes isomorphic to 
$C(\beta\Gamma)/C(\Gamma)\simeq C(\beta\Gamma\backslash\Gamma)$.\\
We thus have
\begin{prop}
Let $A$ be a $C^*$ algebra and let $A^+=A+1{\bf {\sf C}}$ denote the smallest possible
unitization and ${\cal M}(A)$ the multiplier algebra. Suppose the classical
phase-space of $A$ is $\Gamma$ then
\begin{eqnarray*}
    \Gamma(A^+) &\simeq& \alpha \Gamma \mbox{  (one-point compactification)}\\
    \Gamma({\cal M}(A)) &\simeq & \beta \Gamma \mbox{  (Stone-\v{C}ech
    compactification)}
\end{eqnarray*}
\end{prop}
We are now through, ${\cal M}({\cal K}) = {\cal B}$, and as we mentioned, any
non-abelian $C^*$-algebra sits as a subalgebra inside ${\cal B}({\cal H})$ 
for some Hilbert space $\cal H$.\\
With the relationship between unitizations and compactification clarified we 
can go on to extensions. We say that $B$ is an {\em extension} of $A$ by $C$ if
\begin{displaymath}
    0\rightarrow A \stackrel{\alpha}{\rightarrow} B
    \stackrel{\beta}{\rightarrow} C \rightarrow 0
\end{displaymath}
is exact.\\
Now any morphism $0\rightarrow A\rightarrow B$ induces a unique morphism
$B\rightarrow {\cal M}(A)$, in fact we have the following commutative diagram
\begin{displaymath}
    \begin{array}{ccccccccc}
    0 & \rightarrow & A & \stackrel{\alpha}{\rightarrow} & B &
    \stackrel{\beta}{\rightarrow} & C & \rightarrow  & 0\\
      & & \left\| \begin{array}{c}~\\ ~\\ ~\end{array}\right. & &
    \left\downarrow\begin{array}{c}~\\ \sigma\\ ~\end{array}\right.&
   & \left\downarrow\begin{array}{c}~\\ \tau\\ ~\end{array}\right. & &\\
    0 & \rightarrow & A & \rightarrow & {\cal M}(A) & \rightarrow &
    {\cal M}(A)/A & \rightarrow & 0
    \end{array}
\end{displaymath}
the morphism $\tau$ is known as the {\em Busby invariant}, it characterizes the
extension and is unique \cite{Kth}. We suppose we know the classical 
phase-spaces of $A$ and $C$
and we want to find it for the larger algebra $B\subseteq A\oplus C$. It turns
out, \cite{Kth}, that $B$ can be constructed from $\tau$ and $A$ in the following
way
\begin{equation}
    B \simeq \{a\oplus c\in {\cal M}(A)\oplus C~|~ \pi(a)=\tau(c)\}
\end{equation}
where $\pi$ is the canonical quotient map ${\cal M}(A)\rightarrow {\cal M}(A)
/A$. We say that $B$ is the {\em pullback} of ${\cal M}(A)/A$ along $\pi$ and
$\tau$. This implies that $C(\Gamma_B)$ is a kind of ``diagonal'' subspace of
$C(\beta\Gamma_A)\oplus C(\Gamma_C)$, namely:
\begin{prop}
If $A,B,C$ are $C^*$-algebras and if $B$ is an extension of $A$ by $C$ then
\begin{equation}
    C(\Gamma_B) \simeq \{f\oplus g\in C(\beta\Gamma_A)\oplus C(\Gamma_C) ~|~
    \tilde{\pi}(f) = \tilde{\tau}(g)\in C(\beta\Gamma_A\backslash\Gamma_A)\}
\end{equation}
\end{prop}
In this way we are able to construct the classical phase-space of an extension
from its Busby invariant $\tau$ and the classical phase-spaces of the other
algebras. We see e.g. that $C(\Gamma_A)$ has codimension one when $C$ is 
an abelian $C^*$-algebra.\\
Admittedly, the WWM-formalism put forward in this paper is rather formal as far
as $C^*$-algebras are concerned; we were only able to show how in principle one
could construct classical phase-spaces, and we saw that $\Gamma_\infty$, the
classical phase-space of the algebra of compact operators, could be expressed 
as a direct limit of $\Gamma_n=\Gamma(gl_n)$. We have not given explicit
constructions for other $C^*$-algebras though. The next natural step will be to
study specific $C^*$-algebras, e.g. the {\em irrational rotation algebras} 
$A_\theta$, which correspond closely to the Heisenberg algebra, the {\em
Toeplitz algebra}
(generated by the shift-operator), which can be seen as a kind of limit of
$so_l$ or $su_l$, it's generalization the so-called {\em Cuntz algebras} 
and so on. This will be sketched in the next paragraph.

\section{Examples of $C^*$-Algebras}
We will begin with algebras generated by shift operators. First of all, we 
will consider the Hilbert space $l^2({\bf {\sf Z}})$, i.e. the space of all 
square-summable sequences of complex numbers with the set of integers as their
index set. An important operator on this space is the {\em bilateral shift}
\begin{equation}
	S|n\rangle = |n+1\rangle
\end{equation}
where $\{|n\rangle\}, n\in{\bf {\sf Z}}$ denotes an orthonormal basis. The adjoint
operator $S^*$ similarly satisfies
\begin{equation}
	S^*|n\rangle = |n-1\rangle
\end{equation}
and we see that $S$ is unitary. We can form the $C^*$-algebra $A=C^*(S)$ 
generated by $S$ (and thus also including $S^*$). Clearly $A$ is abelian
and hence isomorphic to ${\bf {\sf C}}[[X,\bar{X}]]$, i.e. $\Gamma={\bf {\sf C}}$. A much more
interesting case comes about when we consider not the integers but only the
natural numbers ${\bf {\sf N}}$ as index set. We then get the {\em unilateral shift}, 
which is only an isometry: $S^*S=1$ but $SS^*\neq 1$, in fact $SS^* = 
(1-\delta_{n1}) = 1-P_1$, where $P_1$ is the projection unto $|1\rangle$, i.e.
\begin{equation}
	\left[ S, S^*\right] = P_1
\end{equation}
The corresponding $C^*$-algebra is known as the {\em Toeplitz algebra} and 
will be denoted by ${\cal T}$. This algebra is one of the most 
well-studied an important $C^*$-algebras. It can also be seen as an extension
of $\cal K$, the compact operators, by $C(S^1)$, the abelian $C^*$-algebra
of continous functions on the circle. Any element in $\cal T$
can be written as $x=\sum_{n,m=0}^\infty x_{nm}S^n(S^*)^m = \sum_{n,m} x_{nm} 
T_{nm}$, where $T_{nm} = S^n S^{*m}$. The commutator of these generators
is easily seen to be
\begin{eqnarray}
	\left[ T_{nm}, T_{n'm'}\right] &=& \theta(n'-m)T_{n+n'-m,m'}
	+\theta(m-n')T_{n,m-n'+m'}-\nonumber\\
	&&\theta(n-m')T_{n+n'-m',m}-\theta(m'-n)T_{n',m+m'-n}+\nonumber\\
	&& \delta_{n'm}T_{nm'}-\delta_{nm'}T_{n'm}
\end{eqnarray}
We note that $\{T_{n0}\}, \{T_{0n}\}$ form two (isomorphic) abelian subalgebras.
Any element of the classical phase-space will then be of the form
\begin{equation}
	\xi(x,y) = \sum_{nm}\xi_{nm}x^ny^m
\end{equation}
with
\begin{equation}
	(\xi_{mn})^\dagger = \overline{\xi_{nm}}
\end{equation}
Hence $\Gamma_{\cal T}$ consists of analytical functions $S^1\times S^1
\rightarrow {\bf {\sf C}}$. The ``translation-operator'' $\Pi$ has the form
\begin{equation}
	\Pi_{\cal T}(\xi) = e^{i\sum_{n=0}^\infty\sum_{m=1}^\infty \xi_{nm} 
	T_{nm}+i\sum_{m=0}^\infty \lambda_m(\xi)T_{m0}}
\end{equation}
The only {\em a priori} restriction on the coefficients $\xi_{nm}$ is that $\xi
\in l_1({\bf {\sf N}}_0\times{\bf {\sf N}})$, the set of absolute summable series indexed by
${\bf {\sf N}}_0\times{\bf {\sf N}}$ with ${\bf {\sf N}}_0 = \{0,1,2,3,...\}$. This can also be interpreted
as functions in $H^1(S^1\times S^1)$, the {\em Hardy space} of absolute
integrable functions $f(x,y)$ such that $f$ vanishes whenever $x,y<0$,
and we finally end up with
\begin{equation}
	\Gamma_{\cal T} = H^1(S^1\times S^1)/H^1(S^1) \simeq 
	\{f\in H^1(S^1\times S^1) ~|~f|_{\rm diag} =0\} \equiv \tilde{H}^1
	(S^1\times S^1)
\end{equation}
The Toeplitz algebra can also be defined in another way, namely as the 
$C^*$-algebra generated by operators of the form $x\mapsto T_\phi x = P(\phi 
x)$ where $\phi\in C(S^1)$ and $P$ is the projection $L^2(S^1)\rightarrow 
H^2(S^1)$, so it is not surprising that the Hardy spaces $H^p$ turn up. We get
$H^1$ and not $H^2$ as we only have a norm and not a sesquilinear form on our
operator algebra (if we could defined a ``Hilbert-Schmidt''-subalgebra, then
it would be isomorphic to $\tilde{H}^2$, and we get the space $S^1\times S^1$ 
and not just $S^1$ because we have to take $S$ and $S^*$ as independent 
quantities, thus giving rise to an underlying two-dimensional space.\\
The Toeplitz algebra is not abelian, so it is not surprising that we get an
infinite dimensional phase-space, which we can then, represent as
a space of functions. The elements in the Toeplitz algebra gets represented
by non-linear functionals in this manner.\\
The next obvious step is the socalled {\em Cuntz algebra}, ${\cal O}_n$, 
spanned by $n$ isometries, $S_i$, subject to
\begin{equation}
	\sum_{i=1}^n S_i S_i^* = 1
\end{equation}
i.e. their range projections $S_iS_i^*$ cover the entire space. By analogy
with the Toeplitz case we get
\begin{equation}
	\Gamma_{{\cal O}_n} = \tilde{H}^1(\underbrace{S^1\times S^1\times...
	\times S^1}_{2n})
\end{equation}
The next important case is $A_\theta$ the {\em rotation algebras}, where 
$\theta\in {\bf {\sf R}}$, these are generated by two unitaries $u,v$ subject to
\begin{equation}
	uv = e^{i 2\pi\theta} vu
\end{equation}
Let $T_{nm} = u^nv^m$ we quickly arrive at the algebra
\begin{eqnarray}
	\hspace{-5mm}\left[T_{mn},T_{m'n'}\right] &=& 
	(\delta_{nm'}e^{-in2\pi\theta} -
	\delta_{n'm}e^{-in'2\pi\theta})T_{m+m',n+n'}+\nonumber\\
	&&\theta(n-m')e^{-im'2\pi\theta}T_{m,n+n'-m'} +
	\theta(m'-n)e^{-in2\pi\theta}T_{m+m'-n,n'}-\nonumber\\
	&&\theta(n'-m)e^{-im2\pi\theta}T_{m',n+n'-m}
	-\theta(m-n')e^{-in'2\pi\theta}T_{m+m'-n',n}
\end{eqnarray}
Here $\theta(n)$ is the Heaviside step function.
We see that, when $\theta$ is a rational number, we can choose $n,m,n',m'$ in
a non-trivial way and still get a vanishing commutator (e.g. $n=m', n'=m$ and 
$n-m$ an even number), whereas for $\theta$ irrational this is not possible.
Thus for $\theta\in{\bf {\sf Q}}$ we can have either a larger maximal abelian subalgebra
or we can imbed $l^1({\bf {\sf Z}})$ in more than two (inequivalent) ways. When the
angle $\theta$ is irrational we get
\begin{equation}
	\Gamma_{A_\theta} = l^1({\bf {\sf Z}}^2)/l^1({\bf {\sf Z}}) = \{ (\xi_{nm})\in l^1({\bf {\sf Z}}^2) ~|~
	x_{nn} = 0\}\equiv\tilde{l}^1({\bf {\sf Z}}^2)
\end{equation}
represented as a space of sequences, or equivalently as a space of functions
\begin{equation}
	\Gamma_{A_\theta} = \tilde{L}^1(S^1\times S^1) \equiv \{f\in L^1
	(S^1\times S^1)~|~ f|_{\rm diag} = 0\}
\end{equation}
Further examples can of course be thought of, but we will stop for now. The
spaces we found are listed in table 3. The reason why we always had $\Gamma$ of
the form ${\cal F}(\Gamma_0)$ where $\cal F$ denotes some class of functions
with $\Gamma_0$ compact (indeed of the form $S^1\times\times...\times S^1$) was
that we always had a finite number of generators.

\section{Outlook: Towards a General Dequantization \& Quantization Procedure}
The method we have been developing in the previous sections constitute a 
general
``dequantization'' mechanism: to a given quantum phase-space we associate a
classical phase-space and we identify the quantum operators with functions on
this space. So far this formalism has been developed for Lie, super-Lie
and quantum-Lie algebras aswell as $C^*$-algebras.\\
If we would like to include non-continous functions, we would have to go to von
Neumann algebras instead, and this would be the next natural step. Let me just
sketch what one should probably do. A {\em weight} on a von Neumann algebra 
$\cal A$ is a linear map $\omega:{\cal A}_+\rightarrow {\bf {\sf R}}_+\cup\{\infty\} = [0,
\infty]$, 
we call it a {\em trace} if $\omega(A^*A)=\omega(AA^*)$.\footnote{We mentioned
the possibility of this more abstract definition already in the section on Lie
algebras, but this is the first time we really do need it. For finite 
dimensional
algebras any trace as defined above is just the usual matrix trace (upto a
constant). A further generalization, suited for K-theoretic analysis, is to
replace the trace by an arbitrary cyclic cocycle.}
Any von Neumann algebra
possesses a trace which is semifinite (i.e. the subset of $\cal A$ given by 
$\omega (|A|) <\infty$ is dense in some specific topology). This should be 
the mapping that replaces the usual trace, and we could define
\begin{equation}
    {\cal A}^p(\omega) \equiv \{A\in{\cal A}~|~\omega(|A|^p)<\infty\}
\end{equation}
We then want a map $\Pi$ such that
\begin{displaymath}
    \Pi: {\cal A}^p(\omega) \rightarrow L^p(\Gamma, d\mu_\omega)
\end{displaymath}
is an isomorphism. Continuing as before we would write
\begin{eqnarray*}
    A_W(\xi) &=& \omega(\Pi(\xi)A)\\
    A &=& \int \Pi(\xi)A_W(\xi)d\mu_\omega
\end{eqnarray*}
assuming that we can still use the same $\Pi(\xi)$ in both directions. The
mapping $A\leftrightarrow A_W$ is then also denoted by $\Pi$ as before.\\
The elements of $\cal A$ which do not belong to any of the subspaces ${\cal
A}^p$ would then, by extension of $\Pi$, be mapped into measurable, but not
absolutely integrable functions (i.e. in none of the $L^p$-spaces), i.e.
\begin{displaymath}
    \Pi:{\cal A} \rightarrow M(\Gamma, d\mu_\omega)
\end{displaymath}
where $M(\Gamma,d\mu)$ denotes the set of measurable functions on $\Gamma$. 
We can extend $\Pi$
to all of $\cal A$ by using its semifiniteness, and assuming $\Pi$ to be
continous in some given topology. We know, formally at least, that we {\em 
can} extend our WWM-formalism to von Neumann algebras aswell, as these are,
by definition, subalgebras of ${\cal B}(H)$ for some Hilbert space $H$, i.e.
they lie inside some $C^*$-algebra. Similarly, given any $C^*$-algebra $A$ we
can use the GNS-construction to obtain an isomorphism $\pi$ of $A$ unto a
subalgebra of ${\cal B}(H)$ for some (in general huge) Hilbert space $H$, the
algebra $B=\pi(A)''$ will then be a von Neumann algebra containing $A$, where
$A''$ denotes the double commutant of an algebra (i.e. set of all elements 
which commutes with any element of ${\cal B}(H)$ commuting with all of $A$).\\
As far as operator algebras are concerned, one might also consider 
``regularizing'' the trace, by replacing it by some cyclic cocycle
cohomologous to it.\\
Another important development would be the inverse of what we have been doing
sofar, namely constructing a general {\em quantization} mechanism, which, given
a symplectic manifold deforms it and yields a non-abelian algebra of 
functions which is isomorphic to an operator algebra. Symbolically:
\begin{displaymath}
    \{\cdot,\cdot\}_{PB} \rightarrow [\cdot,\cdot]_M \rightarrow [\cdot,\cdot]
\end{displaymath}
This would allow us to quantize arbitrary classical theories. Some progress has
been made over the past decades in this direction, it is for instance known that
any symplectic manifolds admits a twisted product \cite{Flato}. In this case 
we should
probably make much more use of the symmetries of the classical phase-space,
finding some way, this restricts the corresponding quantum phase-space's
algebraic structure.\\
An interesting application of this formalism would be to index theorems; as the
WWM-formalism establishes a link between operators and functions, and thus
between algebra, geometry and topology, it ought to be useful in this context.
It also opens up the possibility of characterizing the topology of certain
manifolds by purely algebraic means, and, on the other hand, to give
geometrical/topological interpretations of otherwise purely algebraic concepts.
What could turn out to be particularly useful is the various possible choices of
phase-spaces for the algebras $so(r,s)$, depending on whether one looks upon
them as Lie or Clifford algebras, or, indeed, as deformed algebras, establishing
connections between ordinary manifolds, Grassmann spaces and braided spaces
respectively. Especially for harmonic and/or functional analysis on these
spaces, this relationship could very well prove itself very powerfull.\\ 
As a final comment one
should notice that WWM-quantization might help resolve problems of operator
ordering (each WWM-map defined its own unique operator ordering prescription)
and renormalization. The usual problems with renormalizability stems from the
multiplication of distributions, and this is ill defined for ordinary products,
but might be quite reasonable for twisted products, or by ``regularizing'' by
replacing the trace by a cyclic cocycle cohomologous to it.

\section{Conclusion}
We have seen how we can generalize the Wigner-Weyl-Moyal formalism first to 
the case
where the quantum phase-space is an arbitrary Lie algebra of finite or infinite
dimension. We also saw how to relate the WWM formalism for a loop algebra, ${\bf
g}_{\rm loop}$ or a Kac-Moody algebra $\hat{\bf g}_k$ based on some ordinary,
finite dimensional, semisimple 
Lie algebra $\bf g$ to the WWM formalism of $\bf g$ itself. We were furthermore
able to
treat fermionic degrees of freedom, i.e. anticommutators, and hence to include
super-Lie algebras aswell. Next, it was indicated how deformed Lie
algebras, quantum Lie algebras, could be treated too, and how the WWM formalism
of a q-deformed Lie algebra ${\bf g}_q$, could be related to that of the
original algebra. Some comments were also made on intermediate statistics. As
our standard example we took $su_2$, and we saw how the corresponding classical
phase-space turned out to be $S^2$. Naively, the classical phase-space
corresponding to a Lie algebra of rank $l$ and dimension $n$ is ${\bf {\sf R}}^{n-l}$, 
but we realized that
the non-commutativity of the algebra resulted in a deformation of this vector
space, so in the end, the classical phase-space became only locally isomorphic
to ${\bf {\sf R}}^{n-l}$, i.e. became an $(n-l)$-dimensional real manifold. 
The curvature of
this manifold was a measure of the non-commutativity of the Lie algebra. The
algebra structure induced an addition and a symplectic product on the classical
phase-space, which were deformations of the corresponding operations in the flat
space. We should emphasize that although we have only used Lie algebras over the
field of complex numbers, essentially the same analysis should be
possible to carry out with any base-field, e.g. finite fields, thus giving us
{\em Chevalley algebras}, or even just division rings (the quarternions, for 
instance). Some
simplification do occur in our case, though, as ${\bf {\sf C}}$ is algebraically closed.\\
Carried over into the realm of $C^*$-algebras the WWM-formalism
provides us with a kind of non-commutative Gel'fand theorem, which differs from
the usual Gel'fand theorem in the abelian case, though. We also speculated about
how to extend the scheme to include also von Neumann algebras. For reasons of
space, we did not discuss the properties of the corresponding Wigner functions,
this has to be left for future research.

\subsection*{Acknowledgements}
A short version of this paper was presented at the Fourth Wigner Symposium, 
Guadalajara, August 1995, and I'm very greatful for the discussion with the 
participants of that symposium, especially professors Kasperkovitz and 
Schroeck. I am also indebted to professor Dahl for discussions during the 
early phases of this work.

\newpage

\begin{table}[ht]
\centering
\begin{tabular}{|c|c||c|c|}\hline
space & & algebra &\\ \hline
plane & ${\bf {\sf R}}^2$ & $[e,f]=h\qquad [e,h]=[f,h]=0$ & $h_1$\\
cylinder & ${\bf {\sf R}}\times S^1$ & $[e,f]=[h,f]=0\qquad[h,e]=e$ &\\
torus & $S^1\times S^1$ & $[e,f]=0~~[h,e]=ae~~[h,f]=-bf$ &\\
sphere & $S^2$ & $[e,f]=h~~[h,e]=e~~[h,f]=-f$ & $su_2=so_3=sl_2$\\
hyperboloid & $S^{1,1}$ & $[e,f]=-h~~[h,e]=e~~[h,f]=-f$ & $su_{1,1}=so_{2,1} =
sl_{1,1}$\\ \hline
\end{tabular}
\caption{Some particularly simple two-dimensional manifolds and their 
corresponding Lie algebras.}
\end{table}

\begin{table}[h]
\centering
\begin{tabular}{|c|c|c|c|c|c|c|}\hline
        & $e$   & $g_1$ & $g_2$ & $g_3$ & $g_4$ & $g_5$\\ \hline
  $e$   & $e$   & $g_1$ & $g_2$ & $g_3$ & $g_4$ & $g_5$\\
  $g_1$ & $g_1$ & $g_4$ & $g_3$ & $g_5$ & $e$   & $g_2$\\
  $g_2$ & $g_2$ & $g_5$ & $e$   & $g_4$ & $g_3$ & $g_1$\\
  $g_3$ & $g_3$ & $g_2$ & $g_1$ & $e$   & $g_5$ & $g_4$\\ 
  $g_4$ & $g_4$ & $e$   & $g_5$ & $g_2$ & $g_1$ & $g_3$\\
  $g_5$ & $g_5$ & $g_3$ & $g_4$ & $g_1$ & $g_2$ & $e$\\ \hline
\end{tabular}
\caption{The multiplication table of $G=S_3$}
\end{table}

\begin{table}[h]
\centering
\begin{tabular}{|c|c|}\hline
space & $C^*$-algebra \\ \hline
${\bf {\sf C}}[[X,\bar{X}]]$ & bilateral shift\\
$\tilde{H}^1(S^1\times S^1)$ & unilateral shift/Toeplitz algebra\\
$\tilde{H}^1(\underbrace{S^1\times...\times S^1}_{2n})$ & Cuntz algebra
${\cal O}_n$\\
$\tilde{L}^1(S^1\times S^1)$ & irrational rotation algebra \\ \hline
\end{tabular}
\caption{The classical phase-spaces $\Gamma$ for a number of $C^*$-algebras.}
\end{table}

\end{document}